\documentclass[12pt,a4paper]{article}
\usepackage{pdflscape}
\usepackage{amsmath,amssymb,amsthm,amsxtra,overpic,bbm,bm,epsfig,subfigure}
\usepackage{mathrsfs}
\usepackage{booktabs}
\usepackage{graphicx}
\usepackage{xcolor}
\usepackage{comment}
\usepackage{epstopdf}
\usepackage{float}
\usepackage{cite}
\usepackage{array}
\usepackage{makecell}
\usepackage{enumitem}
\usepackage{tabularx} 
\usepackage{overpic}
\usepackage{fancyhdr}
\usepackage{float}
\usepackage{rotating}
\usepackage{color}
\makeatletter
\usepackage{slashed,stmaryrd}
\def\thefootnote{\fnsymbol{footnote}}
\usepackage{geometry}
\definecolor{MyDarkBlue}{rgb}{0.1, 0.1, 0.8} 
\definecolor{SBlue}{rgb}{0.2, 0.4, 0.7} 
\definecolor{MyLightBlue}{rgb}{0.22,0.51,0.9}
\definecolor{MyGreen}{rgb}{0.0, 0.5, 0.0}
\definecolor{BrickRed}{rgb}{0.8, 0.25, 0.33}
\usepackage{hyperref}
\hypersetup{colorlinks, citecolor=SBlue,linkcolor=MyDarkBlue, urlcolor=BrickRed}

\begin{document}
\begin{center}
{\Large \bf 
Probing New Physics at Future Tau Neutrino Telescopes
}
\end{center}
\renewcommand{\thefootnote}{\fnsymbol{footnote}}
\vspace{0.05in}
\begin{center}
{
{}~\textbf{Guo-yuan Huang$^{1}$}\footnote{ E-mail: \textcolor{MyDarkBlue}{guoyuan.huang@mpi-hd.mpg.de}},
{}~\textbf{Sudip Jana$^1$}\footnote{ E-mail: \textcolor{MyDarkBlue}{sudip.jana@mpi-hd.mpg.de}},
{}~\textbf{Manfred Lindner$^1$}\footnote{ E-mail: \textcolor{MyDarkBlue}{manfred.lindner@mpi-hd.mpg.de}},
{}~\textbf{Werner Rodejohann$^{1}$}\footnote{ E-mail: \textcolor{MyDarkBlue}{werner.rodejohann@mpi-hd.mpg.de}}
}
\vspace{0.1cm}
{
\\
\em $^1$Max-Planck-Institut f{\"u}r Kernphysik, Saupfercheckweg 1, 69117 Heidelberg, Germany
} 
\end{center}
\renewcommand{\thefootnote}{\arabic{footnote}}
\setcounter{footnote}{0}
\thispagestyle{empty}
\vspace{0.5cm}
\begin{abstract}
	\noindent
We systematically investigate  new physics scenarios  that can modify the interactions between neutrinos and matter at upcoming tau neutrino telescopes, which will test  neutrino-proton collisions with energies $ \gtrsim 45~{\rm TeV}$, and can provide unique insights to the elusive tau neutrino. At such high energy scales, the impact of parton distribution functions of second and third generations of quarks (usually suppressed) can be comparable to the contribution of first generation with small momentum fraction, hence making tau neutrino telescopes an excellent facility to probe new physics associated with second and third families. Among an inclusive set of particle physics models, we identify  new physics scenarios at tree level that can give competitive contributions to the neutrino cross sections while staying within laboratory constraints:  charged/neutral Higgs and  leptoquarks. Our analysis is close to the actual experimental configurations of the telescopes, and we perform a $\chi^2$-analysis on the energy and angular distributions of the tau events. By numerically solving the propagation equations of neutrino and tau fluxes in matter, we obtain the sensitivities of representative upcoming tau neutrino telescopes, GRAND, POEMMA and Trinity, to the charged Higgs and leptoquark models. While each of the experiments  can achieve a sensitivity better than the current collider reaches for certain models, their combination is remarkably complementary in probing the new physics. In particular, the new physics will affect the energy and angular distributions in different ways at those telescopes.
\end{abstract}
\newpage
\setcounter{footnote}{0}
{
  \hypersetup{linkcolor=black}
  \tableofcontents
}
\newpage

\section{Introduction}
\label{sec:intro}


The IceCube observatory has made significant progresses in measuring the ultra-high-energy (UHE) astrophysical neutrino flux~\cite{IceCube:2013low,IceCube:2013cdw,IceCube:2018cha,IceCube:2021rpz,IceCube:2020abv}. As an elusive messenger, neutrinos have been utilized along with cosmic rays, gamma rays and gravitational waves to understand the nature of cosmic accelerators~\cite{LIGOScientific:2017ync,Anchordoqui:2018qom,Aloiso:2018hbl,HESS:2016pst,Amenomori:2019rjd,HAWC:2019tcx,LHAASO}.
There is a guaranteed flux of UHE neutrinos produced by the scattering of cosmic rays with the cosmic photon background, i.e., the cosmogenic  neutrinos~\cite{Berezinsky:1969erk} with typical energy around ${\rm EeV}\equiv 10^9~{\rm GeV}$, associated with the Greisen-Zatsepin-Kuzmin (GZK) cutoff structure~\cite{Greisen:1966jv,Zatsepin:1966jv,Beresinsky:1969qj} in the cosmic ray spectrum.
After the successful observations of extraterrestrial  UHE neutrinos up to PeV energies at IceCube, a campaign of experimental programs has been launched to measure the cosmogenic neutrinos at extreme EeV energy scales. 
A promising class of such observatories is the tau neutrino telescope, which is sensitive to the $\nu^{}_{\tau}$ component in the cosmogenic neutrino flux.

On the one hand, the  neutrinos can point directly towards the cosmic accelerators without being bent by the magnetic field. On the other hand, the precision measurement of UHE neutrinos can benefit our understanding of fundamental particle physics.
For instance, a Glashow resonance event with shower energy around $6~{\rm PeV}$ has recently been observed by IceCube~\cite{IceCube:2021rpz}, which reinforces the Standard Model (SM) of particle physics~\cite{Anchordoqui:2004eb,Hummer:2010ai,Xing:2011zm,Bhattacharya:2011qu,Barger:2012mz,Barger:2014iua,Palladino:2015uoa,Shoemaker:2015qul,Anchordoqui:2016ewn,Kistler:2016ask,Biehl:2016psj,Huang:2019hgs,Bustamante:2020niz,Zhou:2020oym}. 
Furthermore, the IceCube events up to PeV energies have been used to constrain the neutrino cross section for the first time at a center-of-mass (COM) energy as high as $\sim 1~{\rm TeV}$ for the neutrino-proton collision~\cite{Bustamante:2017xuy,IceCube:2017roe,IceCube:2020rnc}.
Besides verifying the SM predictions, UHE neutrino telescopes are also good facilities to probe certain new physics scenarios beyond the SM~\cite{Ackermann:2019cxh,Ahlers:2018mkf}: test of equivalence principle and Lorentz invariance~\cite{Coleman:1997xq,Gonzalez-Garcia:2005ryx,Arguelles:2015dca,Gonzalez-Garcia:2006koj,Mattingly:2009jf,Murase:2009ah,IceCube:2010fyu,Gorham:2012qs,Esmaili:2014ota,Wang:2016lne,Arguelles:2016rkg,Liao:2017yuy,Stecker:2017gdy,IceCube:2017qyp,Zhang:2018otj,Fiorillo:2020gsb,Chianese:2021vkf,Arguelles:2021kjg,IceCube:2021tdn}, unitarity~\cite{Xu:2014via,Ahlers:2018yom,Song:2020nfh,Denton:2021mso}, fifth forces~\cite{Bustamante:2018mzu}, microscopic black holes~\cite{Uehara:2001yk,Alvarez-Muniz:2002snq,Dutta:2002ca,Kowalski:2002gb,Jain:2002kz,Stojkovic:2005fx,Illana:2005pu,Anchordoqui:2006fn,Kisselev:2010zz,Arsene:2013nca,Reynoso:2013jya,Mack:2019bps}, monopoles~\cite{BAIKAL:2007kno,IceCube:2014xnp,IceCube:2015agw,ANTARES:2017qjw,IceCube:2021eye}, neutrino transition magnetic moment~\cite{Coloma:2017ppo,Coloma:2019qqj}, etc.

The recent anomalies, arising in the measurements of the muon anomalous magnetic moment~\cite{Muong-2:2006rrc,Muong-2:2021ojo} and $B$-meson decays~\cite{Aaij:2021vac, Aaij:2017vbb,LHCb:2019hip}, have indicated the existence of new physics that has a preferable coupling to the second (c, s and $\mu$) and third (b and $\tau$) families over the first one. 
While the second family is still  accessible, the third family is difficult to probe in laboratory. 
In this respect, tau neutrino telescopes are naturally sensitive to the new physics lying in the second and third families. With an EeV incoming neutrino, the COM energy of the neutrino-proton scattering is as high as $45~{\rm TeV}$. This is much higher than what can be achieved in laboratory~\footnote{For instance, FASER$\nu$~\cite{FASER:2019dxq} offers an opportunity to measure neutrino scattering with mean beam energies of 600 GeV to 1 TeV, corresponding to a COM energy less than $45~{\rm GeV}$.}.
At very high energy scales, the parton distribution functions (PDFs) of heavy quarks increase rapidly with small momentum fraction~\cite{Hou:2019efy}, making  processes associated with second and third generations of quark partons, which are suppressed by orders of magnitude at LHC, increasingly important for processes at tau neutrino telescopes.

\begin{figure}[t!]
	\vspace{-0.5cm}
	\begin{center}
		\includegraphics[width=0.95\textwidth]{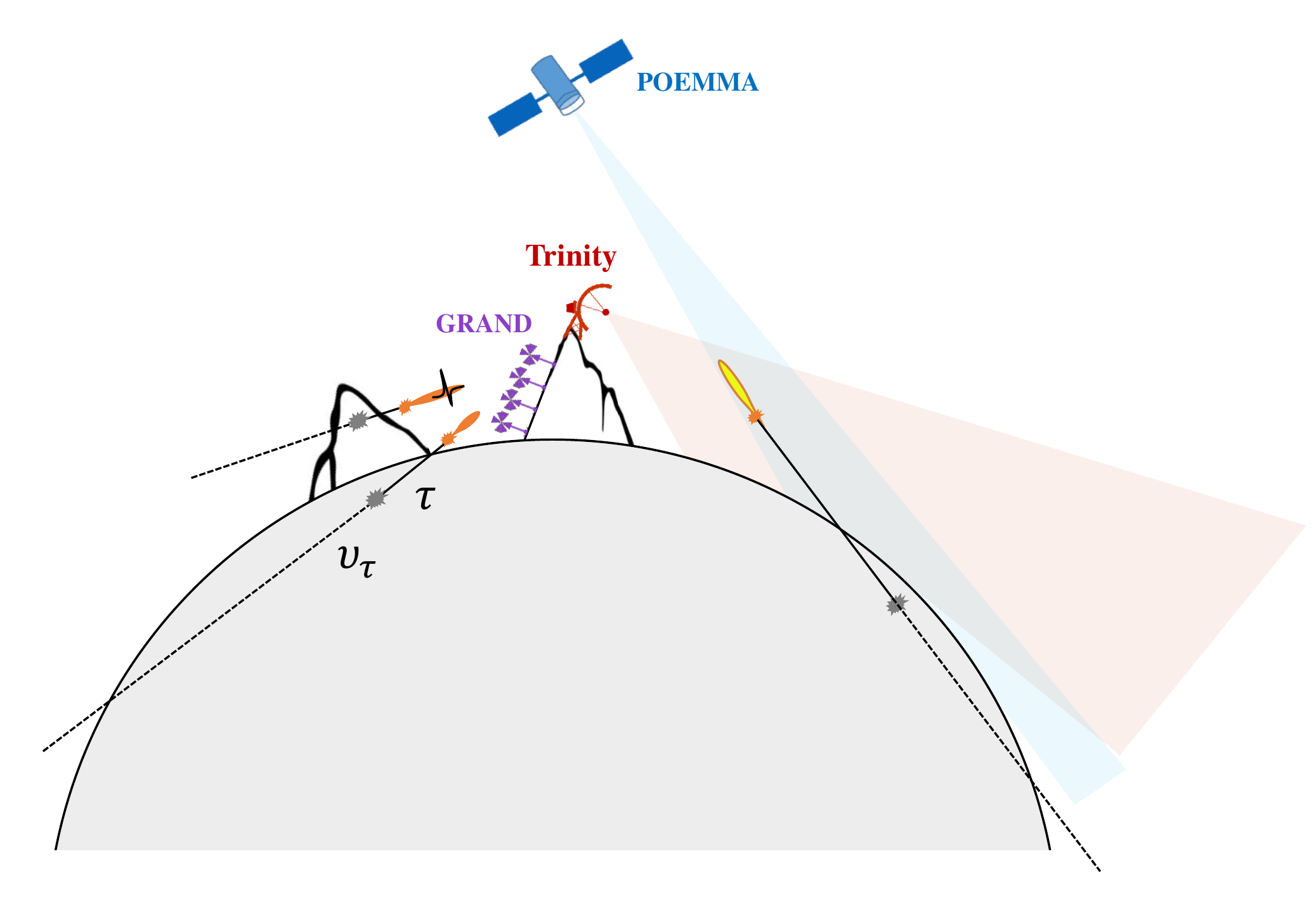}
	\end{center}
	\vspace{-1.2cm}
	\caption{A cartoon of the detection principles of tau neutrino telescopes. Three representative programs with different geography are shown: the mountain-valley telescope GRAND (in purple), the mountain-top telescope Trinity (in red), and the space-borne project POEMMA (in blue). A tau neutrino enters the Earth or mountain, and scatters with matter to produce a tau lepton (the first bang, invisible). The tau lepton eventually decays in the air (the second bang), generating extensive air showers which induce radio signals for GRAND, or Cherenkov light for Trinity and POEMMA. For GRAND, both mountain-penetrating and Earth-skimming neutrinos are observable, but Trinity and POEMMA (in Limb mode) mainly look for  Earth-skimming events near the horizon.}
	\label{fig:demo}
\end{figure}

%
To enhance the effective volume of neutrino interactions, tau neutrino telescopes will be deployed at a high altitude~\cite{Berezinsky:1975zz,Domokos:1997ve,Domokos:1998hz,Capelle:1998zz,Fargion:1999se,Fargion:2000iz,LetessierSelvon:2000kk,Feng:2001ue,Kusenko:2001gj,Bertou:2001vm,Cao:2004sd,Baret:2011zz}, e.g., located on mountains, balloon-, or satellite-borne.
They will monitor the extensive air shower events emerged from the surface of the Earth target.
Since the cosmic rays are shielded by the thick Earth medium, if there is such a shower event, most probably it is produced by a tau decaying in the air~\footnote{The decay length of tau is approximately $c \tau^{}_{\tau} \approx 50~{\rm km}\cdot ({E^{}_{\tau}/{\rm EeV}})$, while that of muon reads $c \tau^{}_{\mu} \approx 4\times 10^8~{\rm km}\cdot ({E^{}_{\mu}/{\rm EeV}})$, which is much larger than the Earth diameter for a typical energy of ${\rm EeV}$. Electrons, on the other hand, will immediately lead to a cascade in the medium after being produced.}, which in turn is generated from a tau neutrino interacting with the Earth.
The extensive air shower events can then be detected in the form of radio waves, Cherenkov light, or fluorescence~\cite{Jelley1958erenkovRI,1953Nature,Greisen1966,Kampert:2012vi,Schroder:2016hrv}. A schematic diagram for the working principles of tau neutrino telescopes is given in Fig.~\ref{fig:demo}, where we have chosen three representative tau neutrino telescopes GRAND~\cite{GRAND:2018iaj,Kotera:2021ca}, Trinity~\cite{Otte:2018uxj,Otte:2019aaf,Wang:2021/M,Brown:2021tf} and POEMMA~\cite{POEMMA:2020ykm}.

\begin{figure}[t!]
	\begin{center}
		\includegraphics[width=1\textwidth]{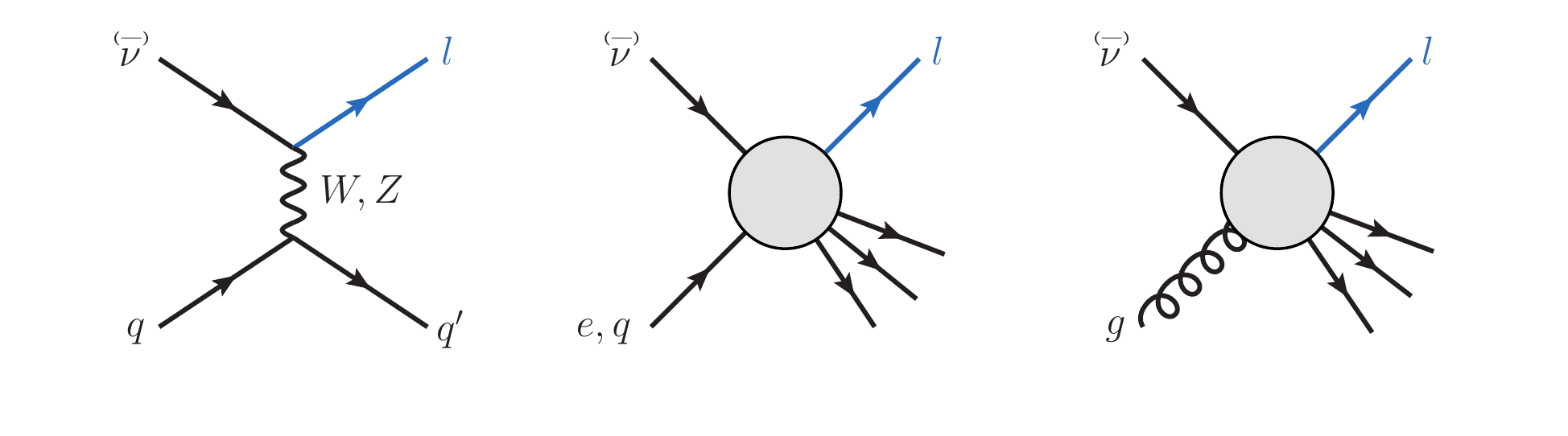}
	\end{center}
	\vspace{-1.5cm}
	\caption{The leading diagrams responsible for neutrino propagation and tau production at tau neutrino telescopes. In the SM, the deep inelastic scattering (the left panel) of both charged-current and neutral-current dominates at our concerned energy scale, i.e.\  $\gtrsim 10~{\rm PeV}$. The possible new physics contributions (middle and right panels) include the neutrino scatterings off electron, as well as quark and gluon partons in the nucleon. Whereas, the process initiated by photon partons is suppressed.}
	\label{fig:SM&NP}
\end{figure}

Given the numerous upcoming programs which search for neutrinos at EeV energies, we intend to systematically investigate the sensitivities of tau neutrino telescopes to the particle physics models modifying neutrino-matter interactions~\cite{Jezo:2014kla,Denton:2020jft,Valera:2021dix,Soto:2021vdc}, with an emphasis on the tau sector.
The new physics scenarios we consider enter into  neutrino scattering processes as in Fig.~\ref{fig:SM&NP}, which directly modifies the interactions between neutrinos and normal matter.
Note that in this work, we do not assume the existence of other long lived particles in the new physics sector, e.g., light sterile neutrinos.

The structure of the rest of the work is organized as follows. In Sec.~\ref{sec:II}, we discuss the strategy to propagate neutrinos and taus in matter at extreme energies. In Sec.~\ref{sec:III}, we introduce the future projects aiming for the detection of GZK neutrinos, and set up the frame for three representative tau neutrino telescopes, i.e., GRAND, Trinity and POEMMA. Sec.~\ref{sec:IV} summarizes the typical particle physics models that can modify the interactions between neutrinos and matter. Charged Higgs and leptoquark (LQ) models are found to have the largest contributions. In Sec.~\ref{sec:V}, we illustrate how the new physics modifies the neutrino interaction and study its consequence at tau neutrino telescopes. The sensitivities of GRAND, Trinity and POEMMA telescopes are explored.
We find these three telescopes are complementary to each other in probing the new physics modifying neutrino interactions, due to their characteristic experimental configurations and in the way the new physics changes the energy and angular distributions of the tau events.
Finally, we make our conclusion in Sec.~\ref{sec:VI}.

\section{Neutrino and Tau Propagation} \label{sec:II}
Neutrinos will be severely attenuated while propagating in matter at EeV energies, around which we expect a bump of the cosmogenic neutrino flux~\cite{Anchordoqui:2007fi,Kotera2010,Ahlers:2012rz,Fang:2013vla,Murase:2015ndr,Ahlers:2018fkn,AlvesBatista:2018zui,Groth:2021bub}. 
To set the scale, the charged-current (CC) cross section of neutrinos in matter is roughly $\sigma^{}_{\rm CC} \approx 10^{-32}~{\rm cm}^2 \cdot (E^{}_{\nu}/{\rm EeV})^{0.363}$~\cite{Gandhi:1995tf,Gandhi:1998ri,Formaggio:2012cpf}, corresponding to a mean free path of $L^{\rm CC}_{\nu}\approx 560~{\rm km} \cdot (E^{}_{\nu}/{\rm EeV})^{-0.363}$ in standard rock with a density of $2.65~{\rm g\cdot cm^{-3}}$. 
In comparison, the Earth diameter is around $12742~{\rm km}$.
There are two competing effects of having a large neutrino cross section at tau neutrino telescopes: 
(i) for CC interactions, one can have a larger conversion rate from $\nu^{}_{\tau}$ to tau and thus collect more events at the detector;
(ii) during propagation the neutrino flux will be more attenuated, which implies an opposite consequence.
These effects allow us to extract the information of neutrino cross sections at EeV energies, and probe possible deviations from the SM caused by new physics.

As for tau at or above EeV energies, after being produced from CC interactions they will suffer from significant energy loss in medium before they decay~\cite{Koehne:2013gpa,Lipari:1991ut,Dutta:2000hh,Alvarez-Muniz:2018owm,Garcia:2020jwr,NuSpaceSim:2021hgs,Safa:2021ghs}. 
Four different processes contribute to the energy loss of tau, i.e., pair production, photonuclear reaction, ionization and bremsstrahlung.
Pair production and photonuclear processes dominate the energy loss at the EeV energy scale.
There have been dedicated codes developed to solve the propagation of neutrinos and taus in matter, to our knowledge e.g.,~\texttt{nuSQuIDS}~\cite{Arguelles:2020hss}, \texttt{NuTauSim}~\cite{Alvarez-Muniz:2018owm}, \texttt{NuPropEarth}~\cite{Garcia:2020jwr}, \texttt{nuPyProp}~\cite{NuSpaceSim:2021hgs}, \texttt{PROPOSAL}~\cite{Koehne:2013gpa} and \texttt{TauRunner}~\cite{Safa:2019ege,Safa:2021ghs}. In this work we have developed our own program to propagate neutrinos and taus.

The task is to solve a coupled set of integro-differential equations describing the neutrino and tau propagation. We summarize the basic strategy below.
The propagation equations can be greatly simplified by taking into account that the neutrino oscillation effect is negligible at energies above a few TeV. 
Including possible new physics contributions 
to the cross section and assuming isoscalar nucleon targets, the propagation of neutrino and tau fluxes is governed by the following equation set~\cite{Lipari:1991ut}:
\begin{align}   \label{eq:nupropa1}
\frac{\mathrm{d}}{\mathrm{d} t} \left(\frac{\mathrm{d}\Phi^{}_{\nu} }{ \mathrm{d} E^{}_{\nu}}  \right)  = & -\; N^{}_{\rm A} \rho \left(\sigma^{\rm CC}_{\rm SM} + \sigma^{\rm NC}_{\rm SM} + \sigma^{\rm CC}_{\rm NP} + \sigma^{\rm NC}_{\rm NP} \right)  \frac{\mathrm{d}\Phi^{}_{\nu} }{ \mathrm{d} E^{}_{\nu}}
&& \text{--- }\parbox[c]{2cm}{\small attenuation } \notag\\
  &  + \;  N^{}_{\rm A} \rho \int \mathrm{d} E^{\prime}_{\nu}
  \frac{\mathrm{d}\Phi^{}_{\nu} }{ \mathrm{d} E^{\prime}_{\nu}}  \frac{1}{E^{\prime}_{\nu} } \left.\left(\frac{\mathrm{d}\sigma^{\rm NC}_{\rm SM}}{\mathrm{d} z}+ \frac{\mathrm{d}\sigma^{\rm NC}_{\rm NP}}{\mathrm{d} z} \right)\right|_{z = \frac{E^{}_{\nu}}{E^{\prime}_{\nu}} } 
&& \text{--- }\parbox[c]{3cm}{\small neutral-current regeneration  } \notag\\
  & + \int \mathrm{d} E^{\prime}_{ \tau} \frac{\mathrm{d} \Phi^{}_{\tau}}{\mathrm{d} E^{\prime}_{ \tau}} 
  \frac{1}{E^{\prime}_{\tau}} \frac{\mathrm{d} \Gamma^{}_{\tau} }{ \Gamma^{}_{\tau} \mathrm{d} z} \;,
  && \text{--- }\parbox[c]{3.3cm}{\small tau regeneration } \\ 
  \label{eq:nupropa2}
  \frac{\mathrm{d}}{\mathrm{d} t} \left(\frac{\mathrm{d}\Phi^{}_{\tau} }{ \mathrm{d} E^{}_{\tau}}  \right)  = & - \; \Gamma^{}_{\tau} \frac{\mathrm{d}\Phi^{}_{\tau} }{ \mathrm{d} E^{}_{\tau}} 
  - N^{}_{\rm A} \frac{\rho}{A} \left[ \sigma^{}_{\rm pair} + \sigma^{}_{\rm photo} +  \sigma^{}_{\rm brem} + \sigma^{}_{\rm ion}\right] \frac{\mathrm{d}\Phi^{}_{\tau} }{ \mathrm{d} E^{}_{\tau}}   
  && \text{--- }\parbox[c]{2.7cm}{\small tau decay and hard energy loss} \notag\\
  &  + \;  N^{}_{\rm A} \frac{\rho}{A}\int \mathrm{d} E^{\prime}_{\tau} \frac{\mathrm{d}\Phi^{}_{\tau}}{\mathrm{d} E^{\prime}_{\tau}} \frac{1}{E^{\prime}_{\tau}} 
  \left. \frac{\mathrm{d}\left( \sigma^{}_{\rm pair} + \sigma^{}_{\rm photo} +  \sigma^{}_{\rm brem} + \sigma^{}_{\rm ion} \right)}{\mathrm{d} z} \right|_{z = \frac{E^{}_{\tau}}{E^{\prime}_{\tau}} }
  && \text{--- }\parbox[c]{3cm}{\small regeneration from hard scattering} \notag\\
  &  + \; \rho \frac{\partial}{\partial E^{}_{\tau}} \left[ \left( \beta^{}_{\rm pair} + \beta^{}_{\rm photo} + \beta^{}_{\rm brem} +\beta^{}_{\rm ion} \right) E^{}_{\tau} \frac{\mathrm{d}{\Phi^{}_{\tau}}}{\mathrm{d} E^{}_{\tau}}   \right] 
  && \text{--- }\parbox[c]{3.1cm}{\small continuous energy loss} \notag\\
  &  + \; N^{}_{\rm A} \rho \int \mathrm{d} E^{\prime}_{\nu} \frac{\mathrm{d} \Phi^{}_{\nu}}{\mathrm{d} E^{\prime}_{\nu}} \frac{1}{ E^{\prime}_{ \nu }} \left. \left(\frac{\mathrm{d}\sigma^{\rm CC}_{\rm SM}}{\mathrm{d} z}+ \frac{\mathrm{d}\sigma^{\rm CC}_{\rm NP}}{\mathrm{d} z} \right) \right|_{z = \frac{E^{}_{\tau}}{E^{\prime}_{\nu}} } \;.
   && \text{--- }\parbox[c]{2.5cm}{\small tau conversion from neutrinos} 
\end{align}
Here, $t$ is time or equivalently the distance traveled by neutrinos and taus, $\mathrm{d}\Phi^{}_{\nu} / \mathrm{d} E^{}_{\nu}$ and $\mathrm{d}\Phi^{}_{\tau} / \mathrm{d} E^{}_{\tau}$ are the differential fluxes of neutrinos and tau, respectively, with the derivative with respect to the solid angle $\Omega$ not explicitly shown for convenience, i.e., $\mathrm{d}\Phi^{}_{\nu,\tau} / \mathrm{d} E^{}_{\nu} \equiv \mathrm{d}^2{\Phi}^{0}_{\nu,\tau} / (\mathrm{d} E^{}_{\nu} \mathrm{d} \Omega)$,
the factor $N^{}_{\rm A}$ is the Avogadro constant, $\rho$ is the mass density of matter and $A$ is the mass number of the target atom. 
Furthermore, $\sigma^{\rm CC}_{\rm SM}$ (or $\sigma^{\rm NC}_{\rm SM}$)  and  $\sigma^{\rm CC}_{\rm NP}$ (or $\sigma^{\rm NC}_{\rm NP}$) are the CC (or NC) cross sections of SM and new physics contributions, respectively, $z \equiv 1-y$ represents the fraction of energy going into the final-state lepton with the inelasticity parameter $y$ being the fraction of energy losses, and $\Gamma^{}_{\tau}$ is the tau decay rate. Note that the new physics contributions include possible interference with the SM.
%

\begin{figure}[t!]
	\begin{center}
		\includegraphics[width=0.43\textwidth]{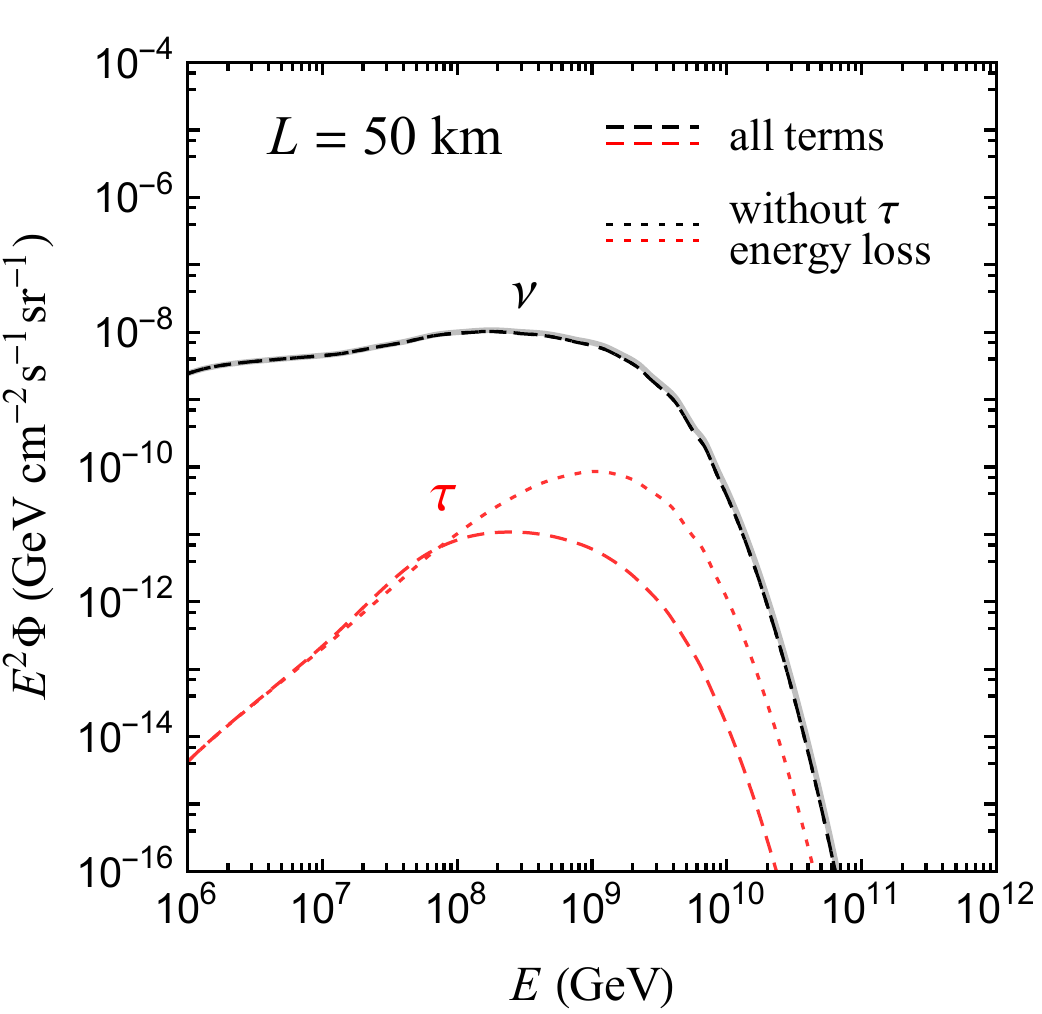}
		\hspace{0.5cm}
		\includegraphics[width=0.43\textwidth]{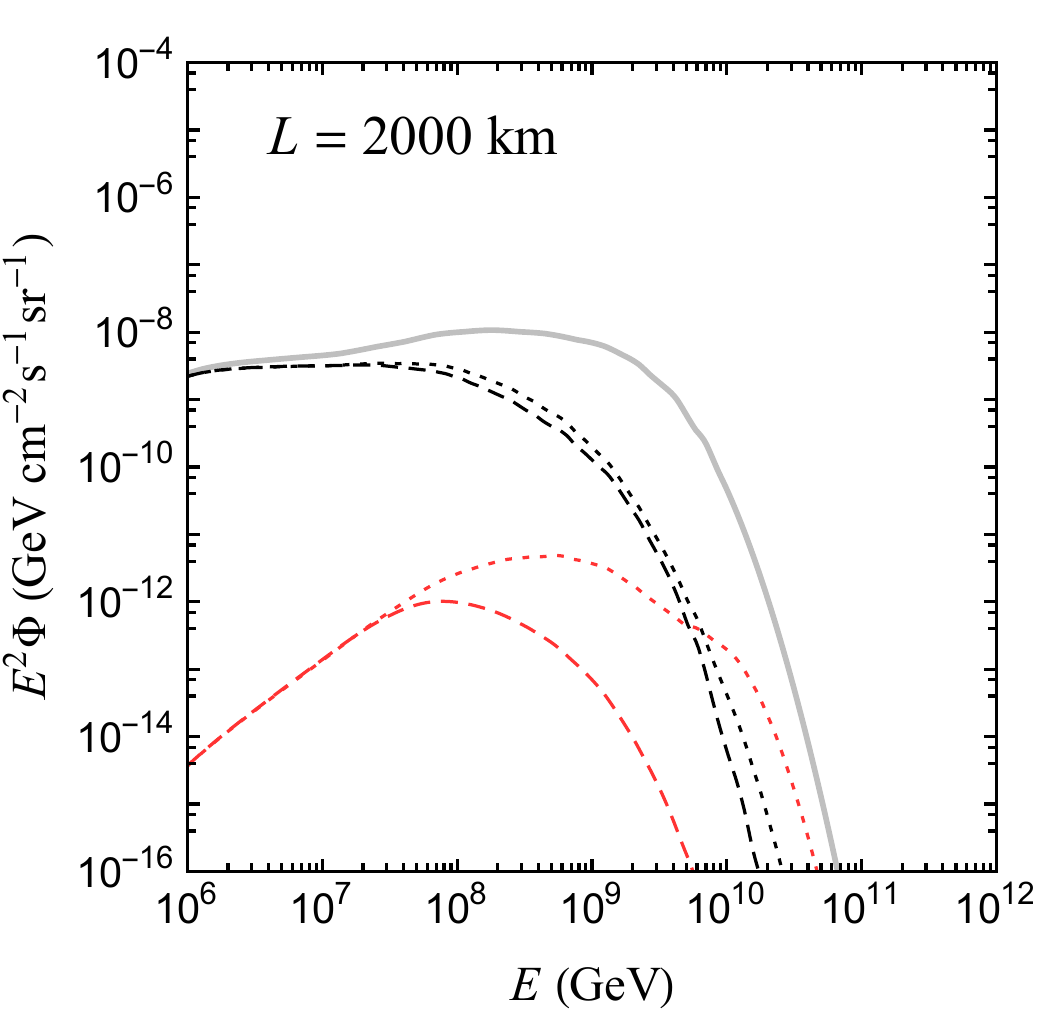}
	\end{center}
	\vspace{-0.3cm}
	\caption{The outputs of neutrino and tau fluxes as a function of the energy, by injecting a cosmogenic neutrino flux (gray curve) into a $50~{\rm km}$ (left-panel) or $2000~{\rm km}$ (right-panel) thick standard rock. The black and red curves stand for the neutrino and tau fluxes, respectively. The complete results including all terms are shown as dashed curves, while those neglecting the tau energy loss are given as dotted curves. Note on the vertical axis  the differential flux $\mathrm{d}^2 \Phi^0 / (\mathrm{d}E \mathrm{d}\Omega)$ has been simplified to a label `$\Phi$' for short.}
	\label{fig:nuprop}
\end{figure}

For other unexplained terms in Eqs.~(\ref{eq:nupropa1}) and (\ref{eq:nupropa2}), $\left\{ \sigma^{}_{\rm pair}, \sigma^{}_{\rm photo}, \sigma^{}_{\rm brem}, \sigma^{}_{\rm ion} \right\}$ stand for the energy loss of tau in matter via pair production, photonuclear, bremsstrahlung and ionization, respectively, and $\left\{ \beta^{}_{\rm pair}, \beta^{}_{\rm photo}, \beta^{}_{\rm brem }, \beta^{}_{\rm ion} \right\}$ are the corresponding parameters for continuous energy loss.
For those processes, the cross section of scattering $\mathrm{d} \sigma / \mathrm{d} y$ increases rapidly and can even be divergent as the inelasticity $y$ goes to zero, while the energy loss rate $\propto y \cdot \mathrm{d} \sigma / \mathrm{d} y$ is always under control. 
For  numerical reasons, the tau energy loss should be separated into  stochastic and continuous contributions~\cite{Lipari:1991ut}.
The stochastic part corresponds to the hard scattering process, where the interaction between tau and medium is dealt with at the cross-section level.
The continuous part collects those effects with very small energy loss per scattering (soft). 
Taking pair production for example, the rate of soft energy loss with $\mathrm{d}E^{}_{\tau} / \mathrm{d} t = - \rho \beta^{}_{\rm pair} E^{}_{\tau}$ reads
\begin{eqnarray}
\beta^{}_{\rm pair} = \frac{N^{}_{\rm A}}{A} \int^{y^{}_{\rm cut}}_{0} \mathrm{d} y \; y \frac{\mathrm{d} \sigma^{}_{\rm pair}}{\mathrm{d} y}\;.
\end{eqnarray}
Here, $y^{}_{\rm cut}$ is a cutoff parameter, above which tau experiences hard scatterings, and below which we can integrate over the inelasticity $(0,y^{}_{\rm cut})$ to obtain the continuous energy loss.
Note that $y^{}_{\rm cut}$ should not be too large, otherwise the stochastic process may not be fully captured.
A reasonable cutoff value would be $y^{}_{\rm cut} \in (0.001,0.01)$~\cite{Koehne:2013gpa}.

To solve the integro-differential equations, we can discretize the momentum space into a number of bins, and translate Eqs.~(\ref{eq:nupropa1}) and (\ref{eq:nupropa2}) into a large set of ordinary differential equations.
In this work, we confine the energy range of interest to be $E \in (10^6,10^{12})~{\rm GeV}$, to which the tau neutrino telescopes are sensitive.
The number of bins should be large enough in order to converge to the actual solution of the differential equations.
We find that a bin number of $300$ is sufficient to have an accurate output of results.

Fig.~\ref{fig:nuprop} demonstrates the solution of neutrino and tau fluxes after a typical cosmogenic neutrino flux from the active galactic nuclei~\cite{Murase:2015ndr} (the gray curve) is injected into a mountain or Earth. The left (right) panel stands for the case that the flux transverses a 50~km (2000~km) thick standard rock with density $\rho = 2.65~{\rm g\cdot cm^{-3}}$. The black and red curves represent the neutrino and tau fluxes, respectively. The dashed curves are the complete results, while
the dotted curves are generated without the tau energy loss.
In the left panel, for $L = 50~{\rm km}$ which corresponds to the typical thickness of a mountain, the neutrino flux experiences a negligible attenuation effect. In comparison, the effect of tau energy losses dominates over the tau decay rate, and becomes significant above $E^{}_{\tau} \sim 10^8~{\rm GeV}$.
In the right panel, the thickness of the standard rock $L = 2000~{\rm km}$ roughly corresponds to a flux emerging from the Earth with an elevation angle of $10^{\circ}$. In this case, the attenuation effect of neutrinos becomes considerable at our concerned energies. The output of tau flux is also reduced compared to the left panel owing to the significantly attenuated neutrino flux in medium.
An ideal choice of the traveled length is roughly the mean free path of neutrinos, where neutrinos have one scattering on average but are not severely depleted. Most of the information about neutrino interactions is contained in these events.

\begin{table}[t!]
	\footnotesize
	\centering
	\caption{Future neutrino telescopes aiming for the detection of cosmogenic neutrinos. In the fourth (fifth) column, we list the sensitive energy (neutrino flavor) of the telescope. The sensitivity to the neutrino flux strength is given in the sixth column, assuming the flux follows a power law spectrum i.e., $E^2_{\nu} \Phi^{}_{\nu} = {\rm constant}$. Note that for simplicity the label `$\Phi^{}_{\nu}$' here stands for the differential spectrum $\mathrm{d}^2 \Phi^0_{\nu} / (\mathrm{d}E \mathrm{d}\Omega)$.
	Abbreviations for `mountain-valley' (Mtn-val), `mountain-top' (Mtn-top), `atmospheric Cherenkov' (Atm-Cher), `fluorescence' (Fluo), `atmospheric radio' (Atm-radio) and `Askaryan effect' (Aska),  have been used. Experiments written in boldface are considered in this work.
We should further emphasize that the sensitivity given here is subject to change by the final experimental design, and the ultimate sensitivity can be  obtained  by properly rescaling from their current proposed exposures.}
	\vspace{0.5cm}
	\begin{tabular*}{\textwidth}{l@{\extracolsep{\fill}}l@{\extracolsep{\fill}} l@{\extracolsep{\fill}}  l@{\extracolsep{\fill}} c@{\extracolsep{\fill}} l@{\extracolsep{\fill}} c@{\extracolsep{\fill}}  }
		\hline
		\hline
		Telescope & Geography & Technique &  {  Energy} & $\nu$ flavor  & { } $E^2_{\nu} \Phi^{}_{\nu}$ & Assumed time \\
		\hline
		\addlinespace[0.8mm]
		EUSO-SPB2~\cite{Adams:2017fjh,Eser:2021H6,Cummings:2020ycz} & Balloon & Atm-Cher, Fluo &  $>10~{\rm EeV}$ & $\nu^{}_{\tau}$ & $2.1 \times 10^{-7}$ & $100~{\rm d}$
\\ \addlinespace[0.8mm]
		PUEO~\cite{Abarr:2020bjd,Vieregg:2021nC} & Balloon & Atm-radio, Aska &  $>0.4~{\rm EeV}$ & $\nu^{}_{\tau}$, $\nu^{}_{e,\mu,\tau}$ & $6.3 \times 10^{-9}$ & $100~{\rm d}$
\\ \addlinespace[0.8mm]
		{\bf POEMMA-Limb}~\cite{POEMMA:2020ykm} & Satellite & Atm-Cher & $>10~{\rm PeV}$ & $\nu^{}_{\tau}$ & $3.2 \times 10^{-9}$ & $5~{\rm yr}$
		\\ \addlinespace[0.8mm]
		POEMMA-Stereo~\cite{POEMMA:2020ykm} & Satellite & Fluo &   $>20~{\rm EeV}$ & $\nu^{}_{\tau}$ & $1.6 \times 10^{-9}$ & $5~{\rm yr}$
		\\ \addlinespace[0.8mm]
		{\bf GRAND}~\cite{GRAND:2018iaj,Kotera:2021ca} & Mtn-val & Atm-radio &  $>50~{\rm PeV}$ & $\nu^{}_{\tau}$ & $1.3\times 10^{-10}$ & $10~{\rm yr}$
\\ \addlinespace[0.8mm]
		TAMBO~\cite{Romero-Wolf:2020pzh} & Mtn-val & Shower particles &  $>3~{\rm PeV}$ & $\nu^{}_{\tau}$ & $4.6 \times 10^{-10}$ & $10~{\rm yr}$
\\ \addlinespace[0.8mm]
		Ashra-NTA~\cite{Ogawa:2021dK} & Mtn-val & Atm-Cher, Fluo &  $>1~{\rm PeV}$ & $\nu^{}_{\tau}$ & $5.5 \times 10^{-10}$ & $10~{\rm yr}$  
\\ \addlinespace[0.8mm]
		{\bf Trinity}~\cite{Otte:2018uxj,Otte:2019aaf,Wang:2021/M,Brown:2021tf} & Mtn-top & Atm-Cher &  $> 1~{\rm PeV}$ & $\nu^{}_{\tau}$ & $5.9 \times 10^{-10}$ & $10~{\rm yr}$
\\ \addlinespace[0.8mm]
		BEACON~\cite{Wissel:2020fav,Wissel:2020sec} & Mtn-top & Atm-radio &  $>10~{\rm PeV}$ & $\nu^{}_{\tau}$ & $1.9 \times 10^{-10}$ & $10~{\rm yr}$
\\ \addlinespace[0.8mm]
		IC-Gen2 Radio~\cite{IceCube-Gen2:2020qha,Hallmann:2021kqk} & In-ice & Aska & $>30~{\rm PeV}$ & $\nu^{}_{e,\mu,\tau}$ & $1.2 \times 10^{-10}$ & $10~{\rm yr}$ 
		\\ \addlinespace[0.8mm]
		RNO-G~\cite{RNO-G:2020rmc,Aguilar:2021uzt} & In-ice & Aska & $>30~{\rm PeV}$  & $\nu^{}_{e,\mu,\tau}$ & $2.4 \times 10^{-9}$ & $10~{\rm yr}$
		\\ \addlinespace[0.8mm]
		ARA~\cite{ARA:2019wcf} & In-ice & Aska & $> 30~{\rm PeV}$ & $\nu^{}_{e,\mu,\tau}$ & $4.3 \times 10^{-9}$ & by $2022$ 
		\\ \addlinespace[0.8mm]
		ARIANNA-200~\cite{Anker:2020lre} & In-ice & Aska & $>10~{\rm PeV}$ & $\nu^{}_{e,\mu,\tau}$  & $1.8 \times 10^{-9}$ & $10~{\rm yr}$
		\\ \addlinespace[0.8mm]
		RET-N~\cite{deVries:2021BA,Prohira:2021vvn,Prohira:2019glh} & In-ice & Radar echo & $>8~{\rm PeV}$ & $\nu^{}_{e,\mu,\tau}$ & $4.0 \times 10^{-10}$ & $5~{\rm yr}$
\\ \addlinespace[0.8mm]
		\hline
		\hline
	\end{tabular*}
	\label{table:nuTel}
\end{table}

\section{Tau Neutrino Telescopes} \label{sec:III}
The water- and ice-based Cherenkov techniques~\cite{Markov:1960vja,Markov:1961tyz} are not optimized for the detection of cosmogenic neutrinos at EeV energies. These include the past programs DUMAND~\cite{DUMAND:1989dxw}, BAIKAL~\cite{BALKANOV2003363} and AMANDA~\cite{ANDRES20001}, the running observatories IceCube~\cite{IceCube:2013low} and ANTARES~\cite{ANTARES:2011hfw}, as well as the future proposals such as IceCube-Gen2~\cite{IceCube-Gen2:2020qha}, Baikal-GVD~\cite{Baikal-GVD:2018isr}, KM3NeT~\cite{KM3Net:2016zxf} and P-ONE~\cite{P-ONE:2020ljt}. 
Even though the neutrino cross section increases with energies, i.e., $\sigma^{}_{\nu} \propto E^{0.363}$, the UHE neutrino flux per decade in energy usually drops faster with a power law spectrum $\mathrm{d}\Phi/\mathrm{d}\log^{}_{10}{E^{}_{\nu}}\propto E^{-1}_{\nu}$.
For those cosmogenic neutrinos, a much larger detection volume is thus required, which can be achieved by placing the detectors at a high altitude. 
It is the hadronic decay of tau in the air from $\nu^{}_{\tau}$ CC interaction with matter that induces the extensive air shower, which subsequently creates detectable signals like radio waves, Cherenkov light or fluorescence. 
The detector and the Earth target together form a huge telescope, with an unprecedented effective volume (or area).
Depending on the geography of the telescope, they can be typically classified into the following categories:
\begin{itemize}
	\item {\it Balloon-borne telescopes.} The ANITA experiment~\cite{ANITA:2018sgj,ANITA:2020gmv} has tested the feasibility of neutrino detection via radio waves from extensive air showers  as well as the Askaryan effect~\cite{Askaryan:1961pfb,ANITA:2006nif} (sensitive to all neutrino flavors). 
	To increase the detection volume, the radio antennas are attached  with a balloon, e.g., floating with a height of $30-40~{\rm km}$ for ANITA. 
	Though no neutrino signal from the Askaryan effect has been detected so far at ANITA~\cite{ANITA:2019wyx}, there are several extensive air shower events~\cite{ANITA:2016vrp,ANITA:2018sgj,ANITA:2020gmv}, which are suspected to be  tau neutrino candidates.
	Two of them from ANITA's first and third flights are anomalous~\cite{ANITA:2016vrp,ANITA:2018sgj}, as their steep incoming angles are in tension with the Standard Model expectation. New physics explanations typically require introducing long-lived degrees of freedom.
	The fourth flight does not see any anomalous events, but instead 
	four neutrino-like events from the near-horizon direction (Earth-skimming) have been detected~\cite{ANITA:2020gmv}. 
	Whether or not these events are truly of neutrino origin remains to be clarified, which is beyond the scope of the present work.

	\item {\it Space-borne telescopes.}
	The detector can also be attached to a satellite in orbit~\cite{Domokos:1997ve,Capelle:1998zz}, and a much larger detection volume can be achieved compared to the balloon-borne experiment.
	The space-borne telescope is mainly sensitive to extensive air showers from Earth-skimming neutrino events. To compensate the light attenuation over large baseline, the detection of atmospheric Cherenkov light or fluorescence is usually preferred.

	\item {\it Mountain telescopes.} For the mountain-valley experiment, as proposed by Refs.~\cite{Fargion:1999se,Fargion:2000iz}, the detector can be placed on one side of a mountain and monitor another mountain over a valley. 
	Furthermore, there are also proposals to place the detector on the top of the mountain, overwatching a thin strip over the horizon, which we will refer to as mountain-top telescope~\cite{Wissel:2020sec} specifically.
	Compared to balloon- and space-borne telescopes, the ground detector arrays are more extensible and easier to maintain.
	Almost all detection techniques (radio waves, Cherenkov light and fluorescence) can be utilized for the mountain telescope.
\end{itemize}

\begin{figure}[t!]
	\begin{center}
		\includegraphics[width=0.43\textwidth]{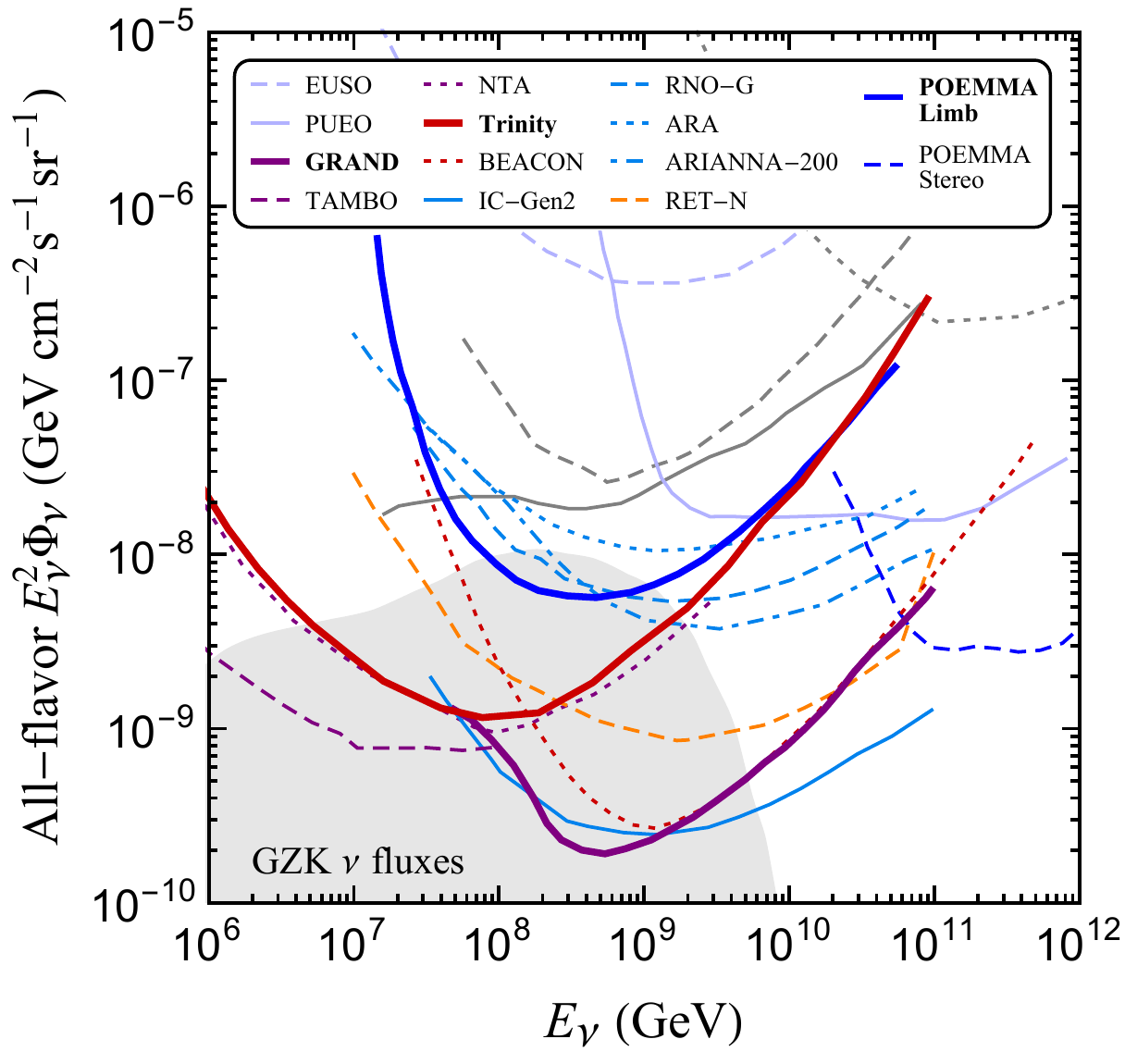}
		\includegraphics[width=0.43\textwidth]{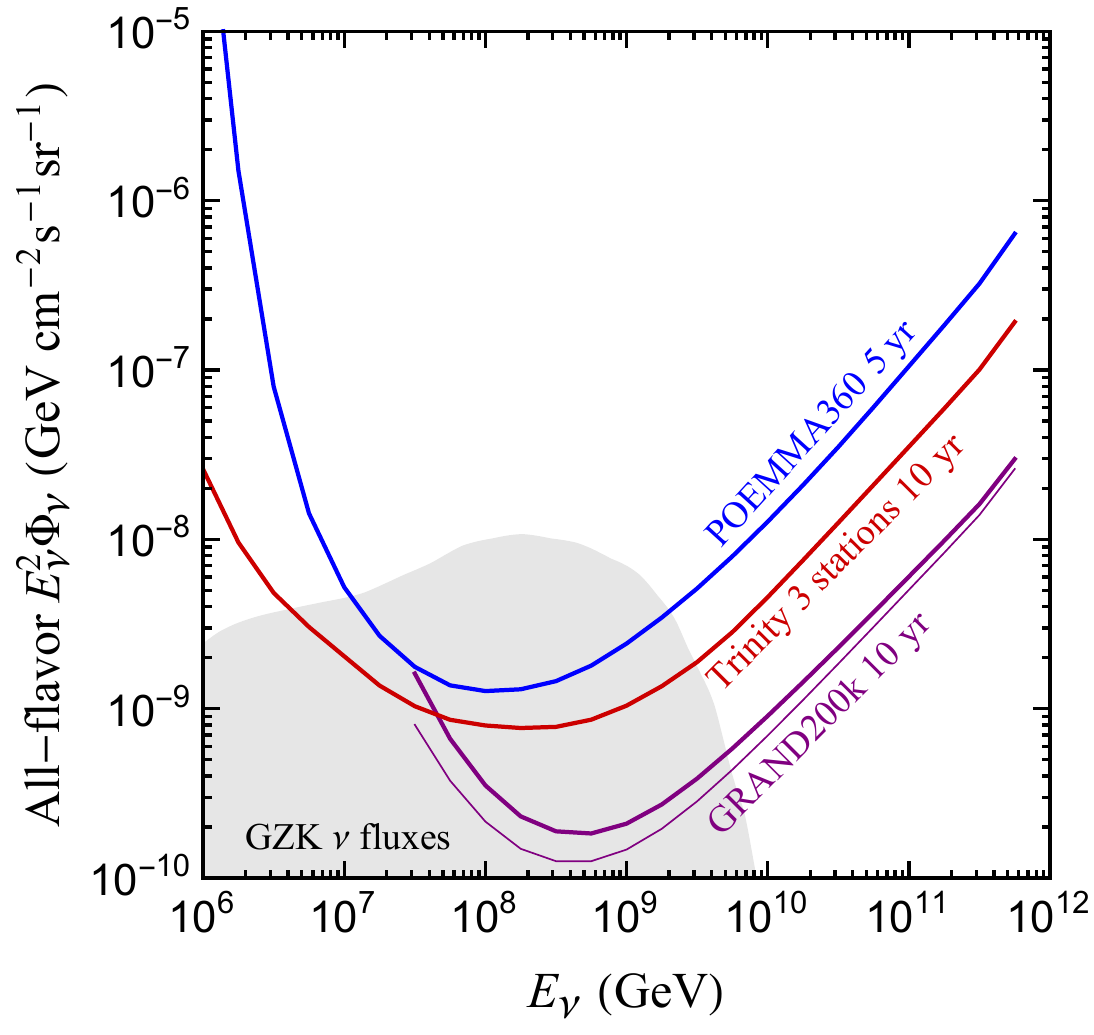}
	\end{center}
	\vspace{-0.3cm}
	\caption{
		{\it Left-panel}: The projected sensitivities of future neutrino telescopes to the all-flavor cosmogenic neutrino flux above PeV energies.
		The details of these experimental projects are collected in Table~\ref{table:nuTel}.
		The sensitivity curves are reproduced or converted from the original proposals~\cite{Adams:2017fjh,Eser:2021H6,Cummings:2020ycz,Abarr:2020bjd,Vieregg:2021nC,POEMMA:2020ykm,GRAND:2018iaj,Kotera:2021ca,Romero-Wolf:2020pzh,Ogawa:2021dK,Otte:2018uxj,Otte:2019aaf,Wang:2021/M,Brown:2021tf,Wissel:2020fav,IceCube-Gen2:2020qha,Hallmann:2021kqk,RNO-G:2020rmc,Aguilar:2021uzt,ARA:2019wcf,Anker:2020lre,Prohira:2021vvn,Prohira:2019glh} for $90\%$ confidence level, i.e., to collect $2.44$ events within a decade of neutrino energy.
		The existing limits from ANITA~\cite{ANITA:2019wyx}, Auger~\cite{PierreAuger:2019ens} and IceCube~\cite{IceCube:2018fhm} experiments are given as dotted, dashed and solid gray curves, respectively.
		The cosmogenic neutrino flux is taken from Ref.~\cite{Murase:2015ndr} for comparison.
		{\it Right-panel}: Our simulated sensitivities to the all-flavor cosmogenic neutrino flux for GRAND200k with 10-year exposure (purple curves), for POEMMA360 with 5-year exposure (blue curve), and for Trinity with 10-year exposure (red curve).
		The thicker (thinner) curve for GRAND is generated assuming the elevation angle of the mountain which hosts the antenna to be $\beta = 3^{\circ}$ ($5^{\circ}$).
		The flavor ratio has been set to $\nu^{}_{e} : \nu^{}_{\mu} : \nu^{}_{\tau} = 1:1:1$ with equal fraction of neutrinos and antineutrinos, and one can simply rescale the curve if a different flavor ratio is chosen.
	}
	\label{fig:fluxSens}
\end{figure}

In Table~\ref{table:nuTel}, we list to the best of our knowledge the existing and proposed telescopes aiming for the detection of cosmogenic neutrinos.
Among these telescopes, the GRAND, Trinity and POEMMA (in Limb mode) experiments represent the mountain-valley, mountain-top and satellite setups, respectively, and have the most outstanding sensitivities to the diffuse tau neutrino flux among similar proposals. We also note that the GRAND proposal has the best diffuse flux sensitivity among these three, and Trinity's sensitivity is slightly better than that of POEMMA in limb mode.
Hence we will take them as three representative prototypes in our later analysis.
Note that we do not consider to explore the potential of all-flavor neutrino telescopes~\cite{Valera:2021dix}, e.g., those with the Askaryan effect, in this work.

To illustrate the detection possibility, in the left panel of Fig.~\ref{fig:fluxSens} we summarize the all-flavor sensitivities of these telescopes to the isotropic diffuse flux of cosmogenic neutrinos, along with the prediction of these neutrinos from the active galactic nuclei~\cite{Murase:2015ndr}.
Those sensitivity curves are reproduced from the corresponding references in Table~\ref{table:nuTel} by requiring the event number over a decade of neutrino energy interval to be $\ln(10) \cdot E^{}_{\nu}\cdot \mathrm{d} N^{}_{}/ \mathrm{d} E^{}_{\nu} = 2.44$, which corresponds to $90\%$ confidence level to observe a positive signal~\cite{Reno:2019jtr}.
For comparison, the observation time has been unified for similar proposals.
To see the current observational status, we recast (with proper rescaling) the existing limits of ANITA~\cite{ANITA:2019wyx}, Auger~\cite{PierreAuger:2019ens} ($\nu^{}_{\tau}$ search) and IceCube~\cite{IceCube:2018fhm} as dotted, dashed and solid gray curves, respectively. None of them are able to provide enough sensitivity to the predicted cosmogenic neutrino flux, yet. But in the future, a larger accumulation time or an experimental upgrade for Auger and IceCube might lead to a discovery of cosmogenic neutrinos. In particular, Auger with a scaling of current exposure by a factor of three will have the potential to observe one event at $90\%$ confidence level, given the cosmogenic flux shown in Fig.~\ref{fig:fluxSens}. However, as we will see in later discussions, to study neutrino interactions a sufficiently larger event number is required, which will be challenging for the current running experiments.
In the right panel, we give the sensitivity results of our simulations, to be discussed in the following.
The notable improvement of POEMMA compared to the left panel should be ascribed to a better experimental configuration~\cite{POEMMA:2020ykm} than the one adopted in the previous estimate~\cite{Reno:2019jtr}. 
We will comment on how accurate our simulation is with respect to the published results when we discuss the experiments in detail.

\subsection{GRAND}
One of the major targets of the GRAND experiment~\cite{GRAND:2018iaj,Kotera:2021ca} is to detect the shower of tau decay initiated by $\nu^{}_{\tau}$ interacting inside the mountain or 
underneath the Earth horizon.
As the extensive air shower propagates in the geomagnetic field a net electric dipole will be developed, which results in strong radio emissions.
The radio antenna array of GRAND will be placed on a slope of the mountain which acts as a large projection screen 
\footnote{This is different from proposals like Ashra-NTA and Trinity, where the detector can be well approximated as point-like. A more complex simulation would be required to obtain the event registered in the large projection surface.}, 
facing towards another mountain which acts as interaction target. The distance between adjacent antennas will be 1 km, such that with an array of 10000 antennas  GRAND10k can cover an inclined surface of $10^4~{\rm km}^2$.
The ultimate stage GRAND200k is to have 20 separate replicates of GRAND10k sites, which can greatly enhance the sensitivity compared to a solo GRAND10k array. With 10-year exposure GRAND200k is able to achieve a world-leading sensitivity $E^2 \mathrm{d}^2 \Phi^0/ (\mathrm{d} E \mathrm{d}\Omega) \sim 10^{-10}~{\rm GeV \cdot cm^{-2}  \cdot sr^{-1} \cdot s^{-1}}$. This is nearly two orders of magnitude beyond the typical floor of cosmogenic neutrino flux, which indicates a capability of collecting $\mathcal{O}(100)$ cosmogenic neutrino events.

An ideal GRAND site should feature both a radio-quiet environment as well as a suitable topography.
A possible location of GRAND10k array is at the Tian Shan Mountain, China, where the simulation of neutrino events has been performed~\cite{GRAND:2018iaj}.
In order to strictly calculate the neutrino event number, the detailed geographical profile near the site is in principle required. 
However, in the present work we consider the following toy setup to capture the main conclusion of GRAND while keeping a phenomenological simplicity.
We assume the antenna array to be uniformly deployed on a $150 \times 66~{\rm km}^2$ mountain slope (inclined by $\beta= 3^\circ$ from the horizontal), opposing to another mountain with the shortest distance in the valley being $10~{\rm km}$. The target mountain has a height of $2.5~{\rm km}$ and a width of $40~{\rm km}$. The length along the valley is set to $150~{\rm km}$, identical to the length of the antenna array. We find that with this toy setup we can closely reproduce the event distributions as well as the sensitivity to tau neutrinos by the official simulation of GRAND.

The antenna array of GRAND10k is wide enough to contain almost all radio pulses from the shower. For instance, the area of a Cherenkov ring imprinted in the antenna screen is around $\pi (l \times 1^{\circ})^2 / \sin{\beta} \sim (7-46)~{\rm km}^2$, where $l = \mathcal{O}{(50)~{\rm km}}$ is the distance from the shower to the  antenna, $1^{\circ}$ is the typical angle of the Cherenkov ring, and $\beta \sim (3^{\circ} - 20^{\circ})$ is the yet-to-be-determined elevation angle of the mountain slope with antennas.
This area is  much smaller than the screen size $\sim 10^4~{\rm km}^2$.
Thus, the simplified methodology assuming a point-like detector, which is much smaller than the scale of a Cherenkov ring, cannot be applied to GRAND.
However, it is a good approximation that an event can be accepted if the neutrino's line of sight intersects with the antenna screen.

The total event number can be obtained by integrating over the neutrino flux coming from different angles:
\begin{eqnarray}
N^{}_{\rm G} = \int \mathrm{d} E^{}_{\tau} \int \mathrm{d} \Omega \int_{S_{\rm 10k}} \mathrm{d}S   \; \frac{\mathrm{d} \Phi^{}_{\tau}}{\mathrm{d} E^{}_{\tau} \mathrm{d} \Omega} \cos{\theta^{}_{\rm tr}} \;  P^{}_{\rm det} \; T \;,
\end{eqnarray}
where $S$ is the area of antenna screen, $\theta^{}_{\rm tr}$ is the angle between the neutrino trajectory and the normal vector of the mountain slope with antennas, and $T=10~{\rm year}$ is the data collection time. The detection probability reads
\begin{eqnarray}
P^{}_{\rm det} = \int \mathrm{d}s \; p^{}_{\rm decay}(E^{}_{\tau}, s)\;  p^{}_{\rm det}(E^{}_{\tau}, \Omega, s) \;,
\end{eqnarray}
where $s$ is the distance traveled by the tau from the mountain or Earth surface before its decay, and the probability density of the decay reads $p^{}_{\rm decay} =\Gamma(E^{}_{\tau})  \mathrm{e}^{-\Gamma(E^{}_{\tau}) s}$ with $\Gamma$ being the tau decay rate. The probability $p^{}_{\rm det}$ is determined by two criteria. First, the tau converted from neutrino should decay and complete the shower development before reaching the antenna array. Second, the induced signal strength at the antenna should exceed the voltage threshold, which has a conservative value $75~{\mu \rm V}$ or a more aggressive one $30~{\mu \rm V}$~\cite{GRAND:2018iaj}.
Namely, 
\begin{eqnarray}
p^{}_{\rm det} = H(s^{}_{\rm 10k} - s) \; H(V^{}_{\rm radio}- V^{\rm min}_{\rm radio}) \;,
\end{eqnarray}
with $H(x)$ being the Heaviside-function. The aggressive threshold $V^{\rm min}_{\rm radio} = 30~{\rm \mu V}$ will be chosen corresponding to the sensitivity curve put by GRAND.
The voltage $V^{}_{\rm radio}$ depending on the detailed antenna response is proportional to the radio flux density which scales as $\propto E^{}_{\tau} /l^2$.
Instead of a simulation of the radio wave production and the response at antennas, we obtain the voltage by using a scaling relation $V^{}_{\rm radio} \sim V^{}_{0} \cdot (E^{}_{\tau}/10^8~{\rm GeV}) \cdot (100~{\rm km}/l)^2$. 
The GRAND results can be approximately reproduced from our toy setup if we take $V^{}_{0} \sim 2~{\rm \mu V}$.
Note that the dependence of antenna's response on the incoming direction is neglected by using this relation. In practice, all these effects should be taken into account with a dedicated simulation of radio waves, which is however beyond the scope of this work.

In Fig.~\ref{fig:fluxSens}, we have shown all-flavor sensitivity of GRAND200k to the diffuse neutrino flux. The thicker (thinner) purple curve stands for the ten-year sensitivity if the elevation angle of the screen mountain slope is chosen to be $\beta = 3^{\circ}$ ($5^{\circ}$) in our toy setup.
With $\beta = 3^{\circ}$ we are able to reproduce the sensitivity curve in Fig.~4 of Ref.~\cite{GRAND:2018iaj} (rescaled to ten years)  at a reasonably close level. 
The event rate benefits from a steeper screen slope, which will possess a larger field of view (FOV) and a better sensitivity to Earth-skimming neutrinos.
Furthermore, the event rate at small neutrino energies, e.g. $E^{}_{\nu} \lesssim 10^8~{\rm GeV}$, is extremely sensitive to the voltage threshold to trigger the antenna, which is subject to the final experimental design.

\subsection{POEMMA}
The POEMMA experiment~\cite{Reno:2019jtr,Anchordoqui:2019omw,POEMMA:2020ykm} consists of two identical satellites in orbit with an altitude of $525~{\rm km}$.
It can operate in two different observation modes: (i)  POEMMA-Stereo, aiming for the detection of cosmic rays or neutrinos above $20~{\rm EeV}$ via the isotropic fluorescence emission of extensive air shower in the atmosphere; (ii)  POEMMA-Limb, for the tau neutrino observation via the Cherenkov light emission of tau decays.
We will focus on the POEMMA-Limb mode, which has a much lower energy threshold, $E^{}_{\nu} \gtrsim 10~{\rm PeV}$, and a better sensitivity at EeV energies than POEMMA-Stereo.

On each POEMMA satellites, there is an optical system which collects and focuses Cherenkov light to the camera. The FOV of the system is $45^{\circ}$, which can be extended to $360^{\circ}$ in azimuth for the POEMMA360 design.
The detection band for Cherenkov light is $200 - 900~{\rm nm}$ for POEMMA. 
To avoid overwhelming backgrounds from the Sun and moonlight, the Cherenkov camera on the satellite can only operate with a $\sim20\%$ duty cycle, which is much smaller than for an experiment based on radio waves. This will limit the effective exposure of such telescopes.
With five years of observation, POEMMA is able to push the sensitivity to cosmogenic neutrino fluxes down to $E^2 \mathrm{d}^2 \Phi^0/ (\mathrm{d} E \mathrm{d}\Omega) \lesssim 10^{-8}~{\rm GeV \cdot cm^{-2}  \cdot sr^{-1} \cdot s^{-1}}$. 

We will closely follow POEMMA's configuration to generate our events\cite{POEMMA:2020ykm}.
For a given initial neutrino fluxes, the event number of POEMMA should be obtained with the formula $N^{}_{\rm P} = \int \int \Phi^{}_{\tau} P^{}_{\rm det}  T \cos{\theta^{}_{\rm tr}} \mathrm{d} S \mathrm{d} \Omega^{}_{\rm tr} $, which can be written more explicitly as
\begin{eqnarray}
N^{}_{\rm P} = \int \mathrm{d} E^{}_{\tau} \int \mathrm{d} \cos{\theta^{}_{\oplus}}  \int \mathrm{d} \cos{\theta^{}_{\rm tr}} \int \mathrm{d}\phi^{}_{\rm tr}  \;  \frac{\mathrm{d} \Phi^{}_{\tau}}{\mathrm{d} E^{}_{\tau} \mathrm{d} \Omega_{\rm tr}} \cos{\theta^{}_{\rm tr}} \; 
2\pi R^{2}_{\oplus} \; P^{}_{\rm det} \; T \;,
\end{eqnarray}
where $\theta^{}_{\oplus}$ is the zenith angle of the tau emergence point with the $z$ axis pointing from the Earth center to the satellite, 
$\theta^{}_{\rm tr}$ is the zenith angle of the tau trajectory but here with $z$ being the vector perpendicular to the Earth surface, and $\phi^{}_{\rm tr}$ is the corresponding azimuth angle.
The output tau flux only depends on the emergence angle $\beta^{}_{\rm tr} = 90^{\circ} - \theta^{}_{\rm tr}$.
The exposure is taken to be
$T = 5~{\rm year} \times 20\%$.
The probability $P^{}_{\rm det}$ that the Cherenkov light from tau decays can be captured by the detector, is determined by
\begin{eqnarray}
P^{}_{\rm det} = \int \mathrm{d}s \; p^{}_{\rm decay}(E^{}_{\tau}, s)\;  p^{}_{\rm det}(E^{}_{\tau}, \theta^{}_{\oplus}, \theta^{}_{\rm tr}, \phi^{}_{\rm tr}, s) \;,
\end{eqnarray}
with $s$ being the distance traveled in the atmosphere before the tau decays.
There are three requirements that a tau decay event will be accepted by the telescope: (i) the satellite camera (approximated as a point) should be within the Cherenkov angle of the shower event; (ii) the tau decay event takes place within the FOV of the telescope camera; (iii) a sufficient number of photoelectrons in the photomultiplier tube (PMT) should be collected.
Hence, the distance-dependent probability $p^{}_{\rm det}$ has the following expression 
\begin{eqnarray} \label{eq:pdetPOEMMA}
p^{}_{\rm det} = H(\theta^{}_{\rm Ch} - \theta) H(s^{}_{\rm FOV} - s) H(N^{}_{\rm PE} - N^{\rm min}_{\rm PE}) \;,
\end{eqnarray}
where we fix the Cherenkov angle as $\theta^{}_{\rm Ch} = 1.5^{\circ}$, and $s^{}_{\rm FOV}$ is the maximal distance within the FOV for a given trajectory. Here, $N^{\rm min}_{\rm PE}$ is the minimum acceptable number of photoelectrons generated by Cherenkov photons registered in the camera, and we fix it as $N^{\rm min}_{\rm PE} = 10$ following POEMMA. The photoelectron number $N^{}_{\rm PE}$ generated by an extensive air shower can be estimated via the relation~\cite{Neronov:2016zou}
\begin{eqnarray}
\frac{\mathrm{d} N^{}_{\rm PE}}{\mathrm{d} \lambda} \approx 3.3\;  \mathrm{e}^{-\tau_{\rm atm} - \tau_{\rm aer}} \left( \frac{E^{}_{\tau}}{10^8~{\rm GeV}} \right) \left( \frac{\lambda}{550~{\rm nm}}\right)^{-2}  \left(\frac{A^{}_{\rm opt}}{4 \pi~{\rm m}^2} \right)  \left(\frac{R^{}_{s}}{10^{3}~{\rm km}} \right)^{-2} \left[ \frac{\epsilon(\lambda)}{0.1}\right] ,
\end{eqnarray}
with $\lambda$ being the photon wavelength in units of nm, $\epsilon(\lambda)$ the quantum efficiency for the photoelectron conversion, $A^{}_{\rm opt}$ the optical area depending on the angle of the incoming photon, $R^{}_{s}$ the distance from the tau decay point to the telescope. 
The overall magnitude is fixed by comparing the distributions to Fig.~17 of Ref.~\cite{POEMMA:2020ykm}.
For POEMMA, the integration should be performed over the frequency band $(200 - 900)~{\rm nm}$. The photon detection efficiency is frequency-dependent, and we will adopt the one for S14520 SiPM array from Fig.~17 of Ref.~\cite{POEMMA:2020ykm}.
The optical area for each POEMMA satellite is $5.71~{\rm m^2}$ on-axis, while for off-axis angles it follows the relation in Fig.~28 of Ref.~\cite{POEMMA:2020ykm}.
The optical depths $\tau^{}_{\rm atm}$ and $\tau^{}_{\rm aer}$, by Rayleigh scattering off the atmosphere and by Mie scattering off aerosols, are estimated following Ref.~\cite{Neronov:2016zou}, assuming the atmosphere height to be $8~{\rm km}$ and the aerosol layer height $1~{\rm km}$.

Now we are ready to compute the $\nu^{}_{\tau}$ events at POEMMA with the tau flux emerged from the Earth surface obtained from the last section.
The sensitivity of POEMMA360 with the $360^{\circ}$ FOV in azimuth to all-flavor diffuse neutrino flux is shown in Fig.~\ref{fig:fluxSens}. Five years of observation with a duty cycle of $20\%$ is able to push the sensitivity curve far beyond the threshold of cosmogenic neutrino flux. 
Our derived sensitivity of POEMMA360 is better than an early evaluation~\cite{Reno:2019jtr}, which should be ascribed to the improved configuration considered in Ref.~\cite{POEMMA:2020ykm}, e.g., a larger optical area and a better quantum efficiency of PMTs in the concerned frequency band.
However, the potential of POEMMA to the diffuse neutrino flux is weaker than the GRAND200k projection around the cosmogenic flux peak $10^8 - 10^9~{\rm GeV}$.

\subsection{Trinity}
Trinity will deploy the imaging atmospheric Cherenkov telescope on the mountaintop with an altitude of $\sim 2~{\rm km}$~\cite{Otte:2018uxj,Otte:2019aaf,Wang:2021/M,Brown:2021tf}.
One may think of Trinity as a ground analogue of  POEMMA, so most of the considerations to derive neutrino events at POEMMA can apply directly to Trinity after modifying some of the experimental parameters.
The complete configuration of Trinity will be made up of three stations, each of which has an imaging system with $360^{\circ}$ field of view in azimuth and $5^{\circ}$ in zenith. 
The horizon of the Trinity telescope with $2~{\rm km}$ height is around $88.56^{\circ}$ from the vertical. A primary investigation of Trinity has found that with $2^{\circ}$ FOV above and $3^{\circ}$ below the horizon, one can achieve an excellent sensitivity for such setups~\cite{Otte:2018uxj}.
The effective collecting area of each Trinity mirror for Cherenkov light is $10~{\rm m}^2$, which can be triggered with a minimum number of $N^{\rm min}_{\rm PE} = 24$ photoelectrons to sufficiently reject the background. Similar to POEMMA, the duty cycle is limited to $20\%$ each year.

Because of the low altitude of Trinity compared to that of POEMMA, Cherenkov photons can be seen by the mirror even far outside the $1.5^{\circ}$ Cherenkov cone.
For instance, a $10^8~{\rm GeV}$ shower that is $135~{\rm km}$ away from the telescope can have sufficient photoelectrons to trigger the mirror, even if the telescope is $10^{\circ}$ away from the shower axis.
To derive the Trinity events, the major difference from POEMMA is in Eq.~(\ref{eq:pdetPOEMMA}):
\begin{eqnarray} \label{eq:}
p^{}_{\rm det} =  H(s^{}_{\rm FOV} - s) H(N^{}_{\rm PE} - N^{\rm min}_{\rm PE}) \;.
\end{eqnarray}
That is, we do not require the telescope to be within the $1.5^{\circ}$ Cherenkov cone. For the photoelectron number $N^{}_{\rm PE}$, we adopt fitted functions in Ref.~\cite{Otte:2018uxj} for a height of  $2~{\rm km}$.
Our calculation of the $10$-year sensitivity of complete Trinity setup is given as the solid red curve in the right panel of Fig.~\ref{fig:fluxSens}, which is reasonably close to the Trinity official result~\cite{Wang:2021/M} in the left panel.

\begin{figure}[t!]
	\begin{center}
		\includegraphics[width=1\textwidth]{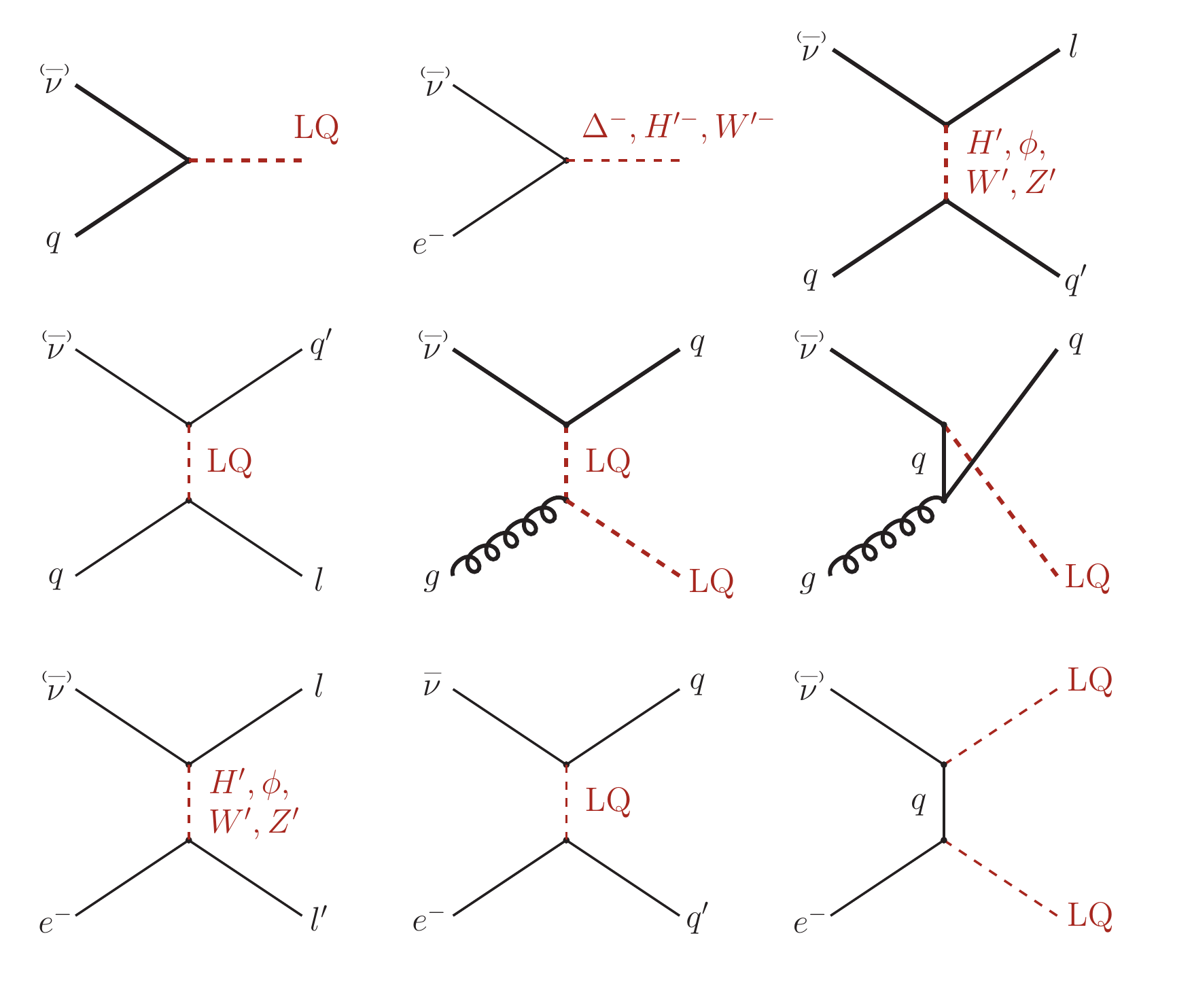}
	\end{center}
	\vspace{-0.3cm}
	\caption{The inclusive contributing diagrams that modify the neutrino-matter interactions as in Fig.~\ref{fig:SM&NP} at tree level.
	Loop-suppressed processes are ignored in this work. The symbol `$l$' in final states stands for the charged lepton or the neutrino. Note that we assume that the baryon number is preserved, $\Delta B = 0$ up to a very high energy scale, and the lepton number can be preserved or violated by two units, i.e. $\Delta L = 0,2$. The new physics diagrams that can induce significant signals at tau neutrino telescopes are highlighted in boldface.}
	\label{fig:feynall}
\end{figure}

\section{Minimal New Physics Scenarios} \label{sec:IV}
In this section we summarize possible new physics contributions that directly modify neutrino-matter interactions at the tree level, which can be probed at tau neutrino telescopes.  
We  maximize the effect to see how much the new physics can contribute to the deviations from the SM.
As has been mentioned, the new physics is involved via the diagrams in Fig.~\ref{fig:SM&NP}, where we must have a neutrino and a matter constituent ($e$, $q$ or $g$) in the initial state and a lepton ($\nu^{}_{e,\mu,\tau}$ or $\tau$) in the final state. The CC production of $e$ and $\mu$ final states is not relevant for tau neutrino telescopes considered here.

Based on the above consideration, we give in Fig.~\ref{fig:feynall} the inclusive new physics cases, where the gauge symmetry and baryon number are always conserved but the lepton number can be possibly violated by two units.
We have shown all  possible contributing diagrams that modify the neutrino-matter interactions at tree level, which include both neutrino-nucleon and neutrino-electron collisions.
One can generalize these scattering processes to a model-independent framework of effective operators by integrating out the heavy degrees of freedom. However, we point out that the leading contribution for some processes  involves resonance production, which is not a trivial task to include in the effective field theory~\cite{Beneke:2004km}.

\subsection{Neutrino-nucleon collision}
The center-of-mass energy of neutrino-nucleon collision reads $\sqrt{s} \approx  \sqrt{2 E^{}_{\nu} M} \approx 43~{\rm TeV}{\sqrt{E^{}_{\nu}/{\rm EeV}}}$ for an incoming neutrino with energy $E^{}_{\nu}$ and nucleon with mass $M \approx 0.94~{\rm GeV}$.
The collision at very high energy scales benefits from the increasing number of sea quarks and gluons in the proton.
For processes relevant for the neutrino-nucleon scattering, we find the following models:
\begin{itemize}
	\item {\it \textbf{Charged Higgs ($H'$) --}}
	There are many SM extensions  such as the two Higgs doublet model \cite{Branco:2011iw}, supersymmetric models, left-right symmetric model \cite{Senjanovic:1975rk}, the type-II seesaw model~\cite{Magg:1980ut,Schechter:1980gr,Cheng:1980qt,Lazarides:1980nt,Mohapatra:1980yp}, radiative neutrino mass models \cite{Babu:2019mfe, Cai:2017jrq, Klein:2019iws, Jana:2019mgj}, axion models \cite{Zhitnitsky:1980tq, Dine:1981rt}   and dark matter models \cite{Abercrombie:2015wmb} possessing  charged Higgs bosons in their particle spectra. 
	A prototypical example can be taken as the Zee model \cite{Zee:1980ai}, which is one of the most well-known neutrino mass models for generating neutrino masses and mixings radiatively at the one-loop level. 
	We consider the following interaction forms
	\begin{eqnarray}
	\mathcal{L}_{H'} \supset y^{q}_{ij} \overline{U}^{}_{i} { D}^{}_{j} H'^+ + y^{\ell}_{\alpha \beta} \overline{\nu^{}_{\rm L}}^{}_{\alpha} E^{}_{{\rm R} \beta}  H'^+ + {\rm h.c.},
	\end{eqnarray}
	where $y^{q}_{ij}$ and $y^{\ell}_{\alpha \beta}$ are the Yukawa coupling constants for quarks and leptons, respectively, $\{i,j \}$ are the quark flavor indices with ${U} \equiv \{\rm u,c,t\}$ and ${D} \equiv \{\rm d,s,b\}$ including both left- and right-handed fields, and $\{\alpha ,\beta \}$ are the lepton flavor indices with $E \equiv \{e,\mu,\tau\}$. In this work we focus on the couplings $y^{q}_{\rm cs}$ and $y^{\ell}_{\tau \tau}$, which are not strictly constrained by laboratory searches.
	
	The combination of $y^{q}_{\rm cs}$ and $y^{\ell}_{\tau \tau}$ will switch on the CC conversion from $\nu^{}_{\tau}$ to tau, i.e., $\nu^{}_{\tau} + {\rm s} ({\rm \bar{c}})\to \tau + {\rm c}({\rm \bar{s}})$. The modification to neutrino cross section with $M^{}_{H'} = 90~{\rm GeV}$ and $y^{q}_{\rm cs} = y^{\ell}_{\tau \tau} =1$ is shown in Fig.~\ref{fig:xsec}.
	Other quark couplings, e.g., $y^{q}_{\rm ud}$, are more severely constrained by the experimental searches. In fact, at the energy scale relevant for tau neutrino telescopes, $y^{q}_{\rm cs}$ can lead to comparable effects as $y^{q}_{\rm ud}$ because of the increasing number of ${\rm c\bar{c}}$ pairs in the nucleon.
	As for other leptonic couplings, $y^{\ell}_{\alpha\beta}$ with $\alpha,\beta = e,\mu$ are not relevant for tau neutrino telescopes.
	Furthermore, $y^{\ell}_{\tau \alpha}$ with $\alpha = e,\mu$ will contribute to the process $\nu^{}_{\tau} \to e,\mu$ during neutrino propagation, resulting in only a depletion in the tau neutrino flux. 
	The effect of $y^{\ell}_{\alpha \tau}$ with $\alpha = e,\mu$ is very similar to $y^{\ell}_{\tau\tau}$ but with the conversion $\nu^{}_{e,\mu} \to \tau$. To be definite, we only switch on $y^{\ell}_{\tau\tau}$ for the later numerical analysis.
	\item {\it  \textbf{Leptoquark (LQ)--}} 
	We demonstrate new physics models with scalar leptoquarks which can be probed at tau neutrino telescopes. Several BSM theories  such as grand unified theories  \cite{Pati:1973uk, Pati:1974yy, Georgi:1974sy, Georgi:1974yf, Georgi:1974my, Fritzsch:1974nn}, radiative neutrino mass models \cite{Babu:2019mfe, Cai:2017jrq, Klein:2019iws, Jana:2019mgj}, technicolor models \cite{Lane:1993wz}, $R$-parity violating supersymmetric models  \cite{Barbier:2004ez}, and dark matter models \cite{Abercrombie:2015wmb} possess these additional colored scalars, i.e., leptoquarks, in their particle spectra. Recently, the leptoquarks have started getting more attention due to their potential to address the $B$-physics anomalies \cite{Aaij:2021vac, Aaij:2017vbb, LHCb:2019hip, Belle:2019gij},
	 muon $g-2$ anomaly \cite{Muong-2:2021ojo, Bigaran:2020jil, Babu:2020hun, Dorsner:2020aaz} and to accommodate sizeable non-standard neutrino interactions \cite{Bischer:2019ttk, Babu:2019mfe}.

	Here we mainly focus on leptoquarks that will induce observable signatures at tau neutrino telescopes. Since it involves a neutrino in the initial state, there are only four leptoquark possibilities which are denoted as $S_1 (3,1,-1/3)$, $S_3 (\Bar{3},3,1/3)$, $R_2(3,2,7/6)$,  $\tilde{R}_2 (3,2,1/6)$ in the literature. Here we are specifying the leptoquarks by their SM quantum numbers $\{\rm SU(3)_C, SU(2)_L, U(1)_Y\}$ and the electric charge is defined as $Q=T_3 + {\rm Y}$. The possible interaction forms for the initial neutrino are $\ell q S_1^{\star}$, $\ell q S_3$, $\ell D^{\rm c} \Tilde{R}_2$, $\ell U^{\rm c} R_2$ at the renormalizable level, with $\ell$ denoting the lepton doublet and $q$ the quark doublet. Note again that we concentrate on  scenarios where there is no extension in the fermionic spectrum in addition to SM fermions. Taking the $S^{}_{1}$ leptoquark as an example, the relevant Yukawa Lagrangian can be expressed as:
	\begin{eqnarray}
	\mathcal{L}_{S_1} \supset -y_{i j}^{\rm L L} \overline{D_{\rm L}^{\rm c i}} S_{1} \nu_{\rm L}^{j}+\left(V^{T} y_{}^{\rm L L}\right)_{i j} \overline{{U}_{\rm L}^{{\rm c}i}} S_{1} E_{\rm L}^{j} +y_{ i j}^{\rm R R} \overline{U_{\rm R}^{{\rm c}i}} S_{1} E_{\rm R}^{j} +\mathrm{h.c.},
	\end{eqnarray}
	where $y^{\rm LL}_{}$ and $y^{\rm RR}_{}$ are Yukawa couplings, and $V$ represents the  Cabibbo-Kobayashi-Maskawa (CKM) mixing matrix.
	For the $S^{}_{1}$ leptoquark, there are two channels contributing to the neutrino scattering: $\nu + D \to {\rm LQ} \to \tau + j \, / \, \nu + j$ and $\nu + g \to {\rm LQ} + j \to  \tau + 2j \, /\, \nu + 2j$.
	Considering collider bounds, we assume the couplings are dominant among the $q={\rm s}$ and $\ell =\tau$ in order to maximize the possible contribution. See the later discussion in this section for more details. The cross section of $S^{}_{1}$ LQ production with $M^{}_{\rm LQ} = 400~{\rm GeV}$ is given in Fig.~\ref{fig:xsec}.
	\item {\it \textbf{Charged gauge boson ($W'$) --}}
	Similar to the charged Higgs, a heavy new charged gauge boson $W'$ can induce similar signatures at tau neutrino telescopes. However, due to the tight constraint from existing experiments~\cite{ATLAS:2019lsy}, i.e., $M^{}_{W'} > 5~{\rm TeV}$, such a heavy $W'$ does not lead to any significant effects at tau neutrino telescopes. There are a few models such as Ref.~\cite{Babu:2018vrl}, where the $W'$ mass can be much lighter ($\sim$ 1.8 TeV) under some broad-width assumption, however we find that this is still not enough to give observable imprint at tau neutrino telescopes; see the left panel of Fig.~\ref{fig:xsec}.
	\item {\it \textbf{Neutral gauge boson ($Z'$) --}}
A new neutral gauge boson $Z'$ exists in many extensions of SM in the gauge sector, such as gauged $L^{}_{\mu}-L^{}_{\tau}$~\cite{He:1990pn,Foot:1990mn,He:1991qd,Baek:2001kca}, $B-L$ models~\cite{Pati:1974yy,Davidson:1978pm,Marshak:1979fm,Wilczek:1979et,Mohapatra:1980qe}, dark sector models~\cite{Bertuzzo:2018ftf, Bertuzzo:2018itn, Berbig:2020wve}, and left-right symmetric model \cite{Senjanovic:1975rk}. To have an observable neutrino-proton scattering process, a $Z'$ should be coupled to both quarks and leptons, for example with the interaction
	\begin{eqnarray}
	\mathcal{L}_{Z'} \supset  \left( g^{U}_{i j} \overline{U^{}_{{\rm L}i}} \gamma^\mu U^{}_{{\rm L}j} +  g^{D}_{i j} \overline{D^{}_{{\rm L}i}} \gamma^\mu D^{}_{{\rm L}j} + g^{\nu}_{i j} \overline{\nu^{}_{{\rm L}i}} \gamma^\mu \nu^{}_{{\rm L}j} + g^{E}_{i j} \overline{E^{}_{{\rm L}i}} \gamma^\mu E^{}_{{\rm L}j} \right) Z^{\prime }_{\mu} \;.
	\end{eqnarray}
	This will contribute to the NC scattering process: $\nu^{}_{\alpha} + q \to \nu^{}_{\beta} + q$. With the same cross section, the effect of NC process is weaker than the CC one, because for the $t$-channel exchange of vector boson the final-state neutrino takes away most of the energy of the  incoming neutrino. This results in an ineffective scattering other than possible mixture among different neutrino flavors.
\begin{figure}[t!]
	\begin{center}
		\includegraphics[width=0.45\textwidth]{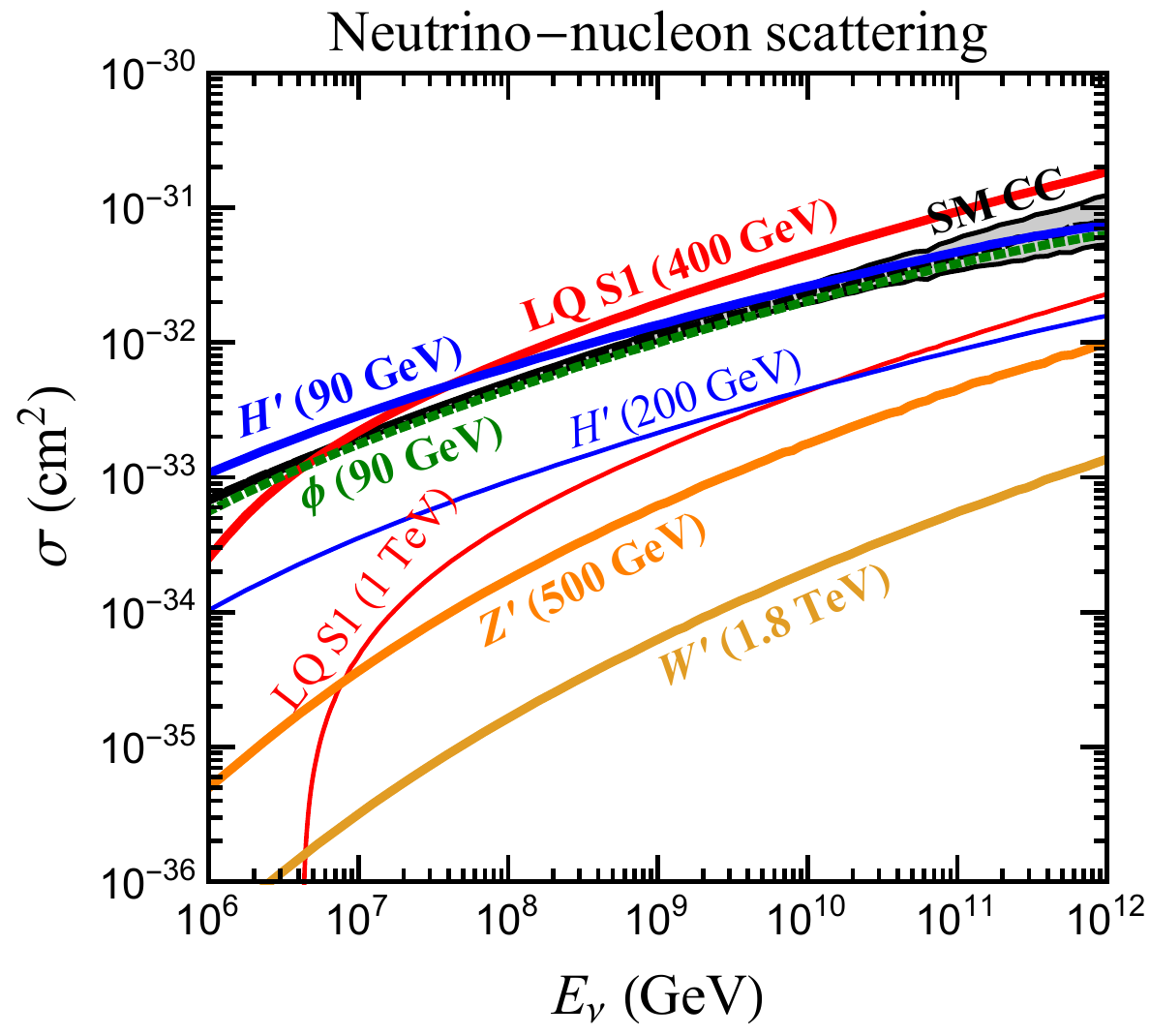}
		\includegraphics[width=0.45\textwidth]{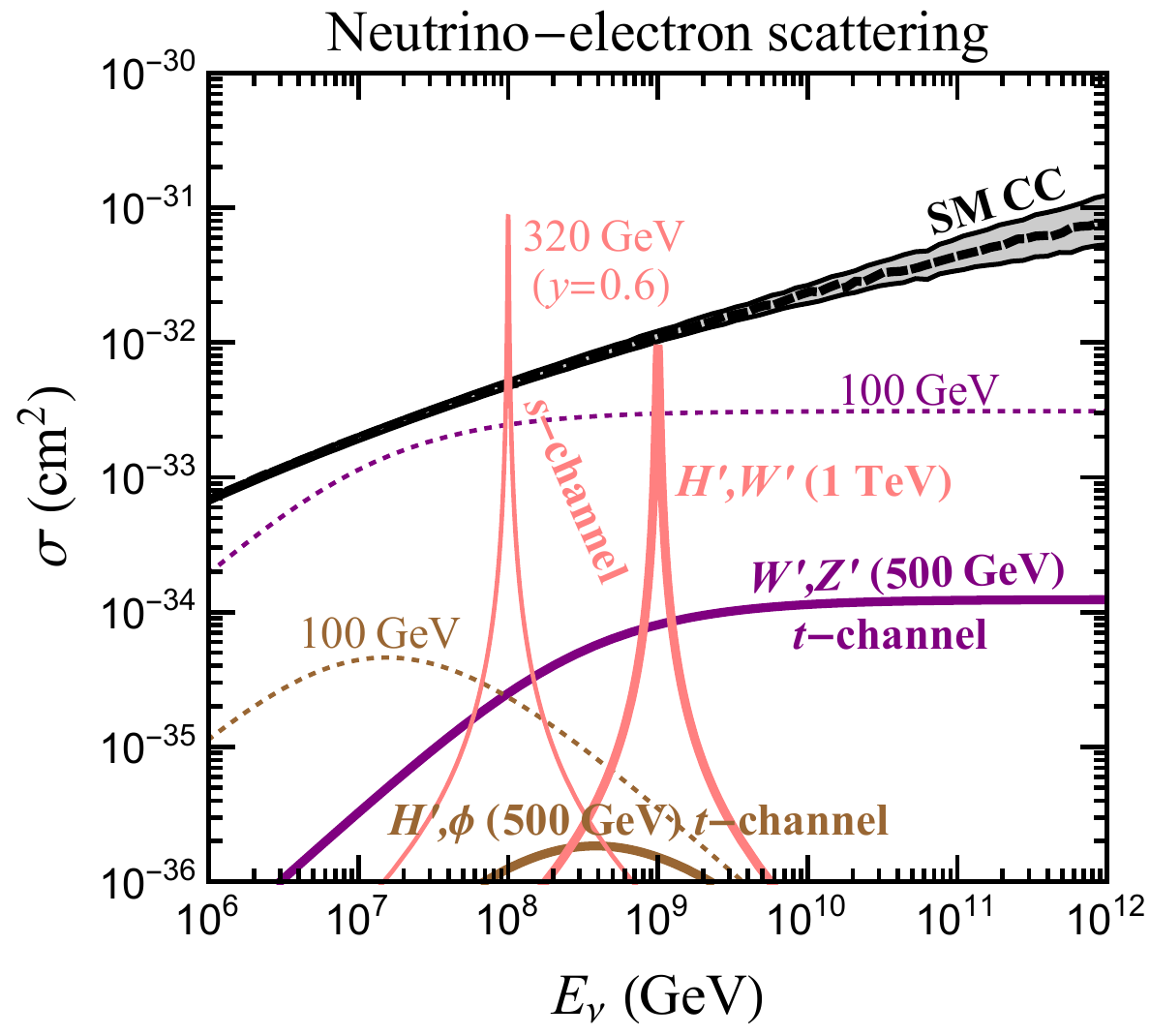}
	\end{center}
	\vspace{-0.3cm}
	\caption{{\it Left-panel}: An illustration of cross sections of the Standard Model CC case (dashed black curve) as well as inclusive new physics scenarios modifying the neutrino-nucleon scattering: leptoquark (in red), charged Higgs $H'$ (in blue), neutral Higgs $\phi$ (in green), charged gauge boson $W'$ (in yellow) and neutral gauge boson $Z'$ (in orange). {\it Right-panel}: The new physics scenarios contributing to the neutrino-electron scattering include: the $s$-channel production of $H'$ and $W'$ (in pink), the $t$-channel exchange of vector bosons $W'$ and $Z'$ (in purple), and the $t$-channel exchange of scalar bosons $H'$ and $\phi$ (in brown).
	The new particle masses are given along the curves, and unless otherwise denoted, all the coupling constants have been taken to be one.
	For the SM CC cross section, the PDF uncertainties are shown as the gray band.}
	\label{fig:xsec}
\end{figure}

	In addition, such $Z'$ coupled to quarks are tightly constrained by collider searches. 
	If $Z'$ talks to quarks of the first family as well as leptons, the lower bound of its mass reads $M^{}_{Z'} \gtrsim 5~{\rm TeV}$~\cite{ATLAS:2019erb}. A minimal scenario is that $Z'$ only talks to the third family, for instance the $U(1)^{(3)}_{\rm B-L}$ model \cite{Babu:2017olk}. The searches of ${\rm bb}\tau\tau$ final states can place a limit $M^{}_{Z'} \gtrsim 100~{\rm GeV}$ for order one couplings~\cite{Elahi:2019drj}.
	However, for an anomaly-free gauge extension consistent with SM~\cite{Babu:2017olk}, $Z'$ will also couple to the first and second generations of quarks via mixing, and the mono-jet searches at LHC set a stringent constraint $M^{}_{Z'} \gtrsim 500~{\rm GeV}$~\cite{Babu:2020nna} with order one couplings.
	In the left panel of Fig.~\ref{fig:xsec}, we show the case for a $Z'$ coupled primarily to bottom quark and tau neutrino with $M^{}_{Z'} = 500~{\rm GeV}$ and an $\mathcal{O}(1)$ coupling.
	\item {\it \textbf{Neutral scalar ($\phi$) -}}
	Various BSM theories predict neutral scalars in their particle spectra, such as the two Higgs doublet model \cite{Branco:2011iw}. Similar to the $Z'$ scenario, a neutral BSM scalar $\phi$ should be coupled to both quarks and leptons to have an observable neutrino-nucleon scattering process. However, similar to $Z'$, if $\phi$  couples to quarks of the first two generations, the coupling is tightly constrained. In addition, if $\phi$ inherits flavor violating couplings, it will give rise to flavor changing neutral-current processes, which is stringently bounded~\cite{Babu:2018uik}.
	This leads us to consider a third-generation-philic scalar scenario, where $\phi$ mostly couples to bottom quarks and tau neutrinos.  This scenario can be constrained by $pp \to {\rm b}\bar{\rm b} + \slashed{E}^{}_{\rm T}$ searches \cite{CMS:2017dcx}. We find that this bound is not very  significant. In the left panel of Fig.~\ref{fig:xsec}, we show the case of $\phi$ coupled to ${\rm b}$ and $\nu_\tau$ with $M^{}_{\phi} = 90~{\rm GeV}$ and an $\mathcal{O}(1)$ coupling.
	Because at large $Q^2$ there are more and more sea quarks (including heavy quarks like c and b) with small momentum fraction, the contribution of a neutral scalar coupled to b quarks is only a factor of two smaller than charged Higgs coupled to s and c quarks. 
\end{itemize}
After considering the above new physics scenarios, we find only the charged/neutral Higgs and leptoquark models can lead to detectable signals of neutrino-nucleon scatterings at tau neutrino telescopes, if the laboratory constraints are taken into account.
In the left panel of Fig.~\ref{fig:xsec}, we depict neutrino-nucleon  cross sections for the pure SM case (gray curve) and inclusive new physics scenarios.
The uncertainties of SM cross section induced by the $1\sigma$ PDF errors computed from the CT18 set~\cite{Hou:2019efy} are also shown for comparison, as the gray band. 
In the following, we shall elaborate on the charged Higgs (the neutral Higgs scenario is similar but weaker) and leptoquark models, which will be further explored in the rest of the work.

\begin{figure}[t!]
	\begin{center}
		\includegraphics[width=0.45\textwidth]{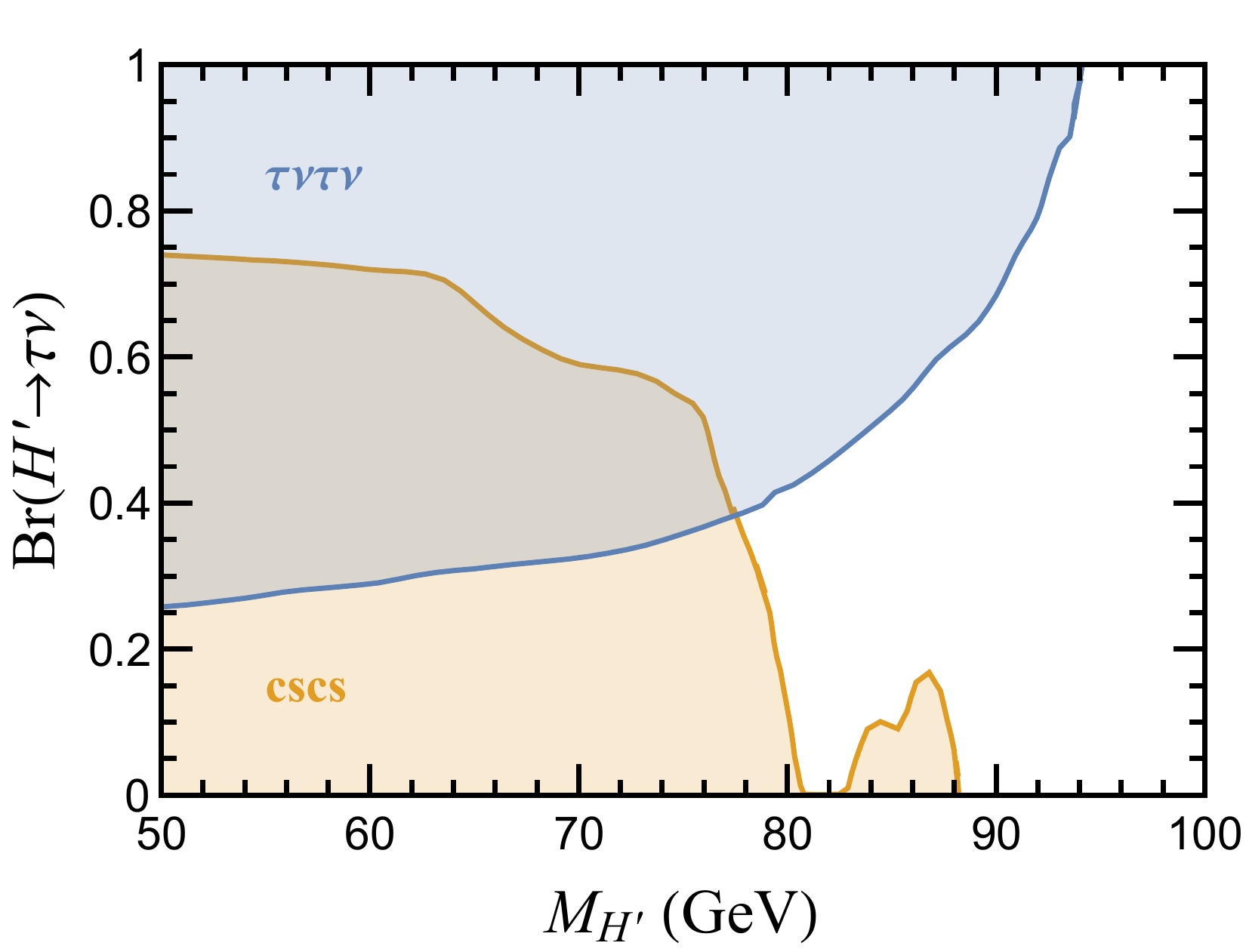}
		\includegraphics[width=0.45\textwidth]{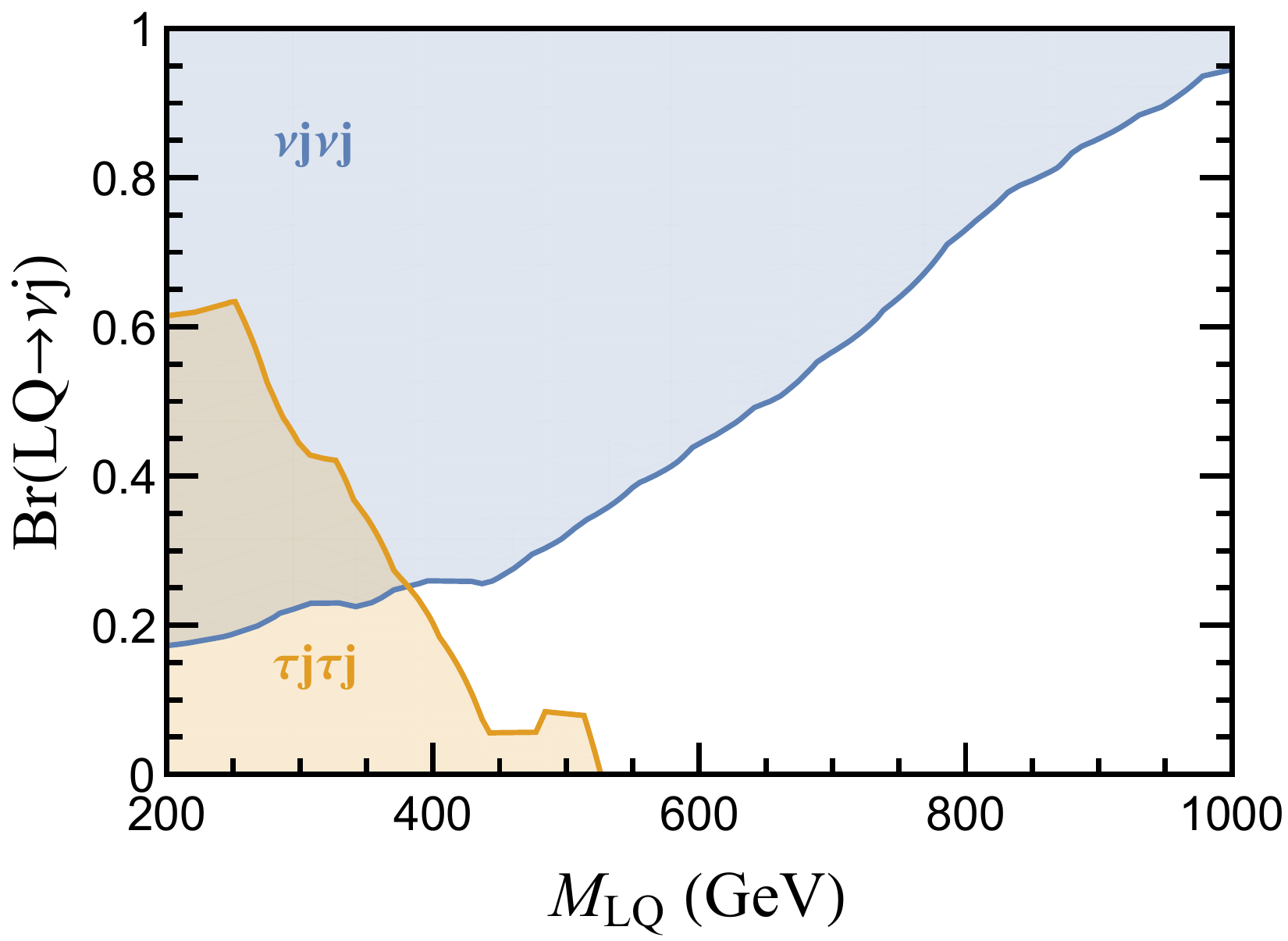}
	\end{center}
	\vspace{-0.3cm}
	\caption{Summary of LEP constraints on the charged Higgs mass from the $s$-channel Drell-Yan process (left panel) and LHC constraints on the leptoquark mass from leptoquark pair production (right panel). The shaded regions are excluded. For the charged Higgs and leptoquark models, we assume the branching ratios to satisfy ${\rm Br}(H' \to \tau \nu )+{\rm Br}(H' \to {\rm c s} ) = 1$ and ${\rm Br}({\rm LQ} \to \nu j)+{\rm Br}({\rm LQ} \to \tau j) = 1$, respectively.}
	\label{CHLQbounds}
\end{figure}

First we briefly discuss the limits on charged Higgs from collider searches. The $s$-channel Drell-Yan process, mediated by either $\gamma$ or $Z$ boson, can pair-produce charged scalars at the LEP experiment. Here we are concentrating on the scenario where the charged scalar mostly couples to $\tau \nu$ and ${\rm cs}$. Hence, the branching ratios of $H'$ decaying to $\nu \tau$ and ${\rm cs}$ modes should add up to one. In the left panel of Fig.~\ref{CHLQbounds}, we show the limits from  charged Higgs searches \cite{ALEPH:2013htx} at LEP  by looking at final state signature ${\rm cs cs}$ or $\tau \nu \tau \nu$. At  LHC, charged Higgs can similarly be produced in pairs via Drell-Yan processes. After being produced in pairs, if the charged Higgs decays back to $\tau \nu $ final state, it will be further constrained from supersymmetric stau searches \cite{CMS:2018yan} because both of them give rise to the same final-state signature in the zero neutralino mass limit. However, we find that in our scenario, where the charged Higgs is emerged from the Zee model similar to  Ref.~\cite{Babu:2018vrl},  the observed cross-section limit is still larger than the theory prediction (for detailed collider analysis, see Ref.~\cite{Babu:2018vrl}). 
Note that there are several dedicated charged Higgs searches \cite{ATLAS:2021upq,ATLAS:2018gfm, CMS:2015yvc}, however in all these scenarios a large $H'{\rm t{b}}$ coupling has been considered. Since in our scenario, the charged Higgs only talks to ${\rm c}\bar{\rm s}$, these bounds are not directly applicable. On the other hand, since the luminosity of ${\rm c}\bar{\rm s}$ quarks is two orders of magnitude lower than that of ${\rm u}\bar{\rm d}$ quarks at LHC \cite{Fuentes-Martin:2020lea}, other search limits are very weak in our scenario.

We continue with the experimental constraints on leptoquark models.
Leptoquarks can be pair produced copiously via $gg$ and $q\bar{q}$ fusion processes at the LHC, and the pair-production rate is uniquely determined by the LQ mass, irrespective of their Yukawa couplings. Leptoquarks can also be singly produced in association with leptons through $s$- and $t$-channel $qg$ fusion processes, and the production rate depends on both LQ mass and LQ Yukawa couplings. 
However, LHC constraints from single-production of LQ are not so severe compared to the pair-production limits \cite{ATLAS:2020dsk, ATLAS:2019qpq} unless the Yukawa couplings to the first and second-generation quarks are too large ($>1$) \cite{Babu:2019mfe,Buonocore:2020erb}.
Other than these direct limits from pair and single production of LQ, there are indirect limits on the mass and Yukawa couplings of  LQs from the dilepton searches at the LHC \cite{ATLAS:2020yat} because the process $pp \to l^+ l^- $ gets significantly modified due to the $t$-channel LQ exchange. However, due to our judiciary choice of Yukawa texture relevant for tau neutrino telescopes, only the process $pp \to \tau^+ \tau^- $  gets modified, keeping the $pp \to e^+ e^-/\mu^+ \mu^- $ processes  almost unaltered. Whereas, the LHC limits are not so strict for $pp \to \tau^+ \tau^- $ signature \cite{Angelescu:2018tyl}. The most stringent limit comes from LQ pair production searches. As we previously mentioned, we are interested in the scenario where LQs mostly decay into $\tau j$ or/and $\nu j$ final state. 

Hence, the searches for the final-state signatures containing two neutrinos and two jets ($\nu \nu j j$) and two tau leptons and two jets ($\tau j \tau j$) impose the best constraints. Although there is a dedicated search for $\nu j\nu j$ final-state signatures  \cite{CMS:2018qqq}, there are no such dedicated searches for $\tau j \tau j$ final states. 
We recast the limits for $\tau j \tau j$ final-state signatures~\cite{Babu:2020hun} from $\tau b \tau \bar{b} $ searches \cite{ATLAS:2019qpq} considering a $b$-jet misidentification rate of 1.5$\%$  as light jets with a $b$-tagging efficiency of 70$\%$ \cite{CMS:2012feb}. The green and purple shaded regions in the right panel of Fig.~\ref{CHLQbounds} depict  the bounds from $\tau j \tau j$ and $\nu j \nu j$ final states. We can see that the former is less stringent than the latter one, and the Yukawa couplings can be chosen in such a way that our LQ can be as light as 400~GeV, making it consistent with all the experimental searches for a branching ratio to $\tau j$ of 80$\%$ and that to $\nu j$ of 20$\%$. 
The branching ratio arrangement requires the LQ Yukawa coupling of $\tau j$ to be stronger than that of $\nu j$, and we see this is natural for $S^{}_{1}$ and $R^{}_{2}$ LQ models. Whereas for $S^{}_{3}$ and $\tilde{R}^{}_{2}$ LQ models, the Yukawa couplings to neutrino and charged leptons are the same, leading to $50\%$ $\nu j$ and $\tau j$ and hence a lower bound $M^{}_{\rm LQ} > 700~{\rm GeV}$; see Appendix for details. The lower limit of LQ mass  can be further reduced by allowing other channels to partly share the branching ratios.
The consequence of leptoquarks on ${\rm km}^3$-scale water- or ice-based Cherenkov detectors has been explored in previous literature~\cite{Anchordoqui:2006wc,Dey:2015eaa,Barger:2013pla,Dutta:2015dka,Mileo:2016zeo,Dev:2016uxj,Dey:2017ede,Chauhan:2017ndd,Becirevic:2018uab}, but the bound and future sensitivity of IceCube have been found to be difficult to exceed the current LHC bounds~\cite{Becirevic:2018uab}.

\subsection{Neutrino-electron collision}

There might be leptophilic forces  which can evade hadron collider limits while affecting the neutrino-electron scattering at tau neutrino telescopes.
The impact can be enhanced by the resonant production of exotic charged particles from neutrino-electron scatterings, e.g., $\nu + e^- \to \Delta^-$ in type-II seesaw scenario or $\overline{\nu} + e^- \to H'^-,W'^-$~\cite{Babu:2019vff,Dey:2020fbx}, similar to the  Glashow resonance in the SM~\cite{Huang:2019hgs}.
For EeV incoming neutrinos, the resonance is at $M^{}_{H'} = \sqrt{s} \approx 1~{\rm TeV}$.
The cross section for the process $\overline{\nu}^{}_{\alpha} + e^- \to H'^- $ reads~\cite{Babu:2019vff}
\begin{eqnarray}
\sigma (s) = 8 \pi \Gamma^2_{H'} \frac{s/M^2_{H'}}{(s-M^2_{H'})^2+(M^{}_{H'}\Gamma^{}_{H'})^2} \;,
\end{eqnarray}
where $\Gamma^{}_{H'} = Y^2 M^{}_{ H'} /(16\pi)$ is the decay width with $Y$ being the coupling strength.
The resonance will be smeared due to the spread of initial neutrino flux and the energy distribution of final states.
Integrating over a decade of neutrino energy around the resonance, we obtain the flux-averaged cross section as $2 \times 10^{-34}~{\rm cm}^2$ for $M^{}_{H'} = 1~{\rm TeV}$ and $Y=1$.
This should be compared to the CC cross section $\sigma^{}_{\rm CC}(1~{\rm EeV}) \approx 10^{-32}~{\rm cm}^2$.
The smallness of resonance enhancement can be ascribed to the relation $\sigma^{}_{\rm res} \propto M^{-2}_{H'}$.
Hence, we cannot gain much from the resonance at the energy scale of cosmogenic neutrinos, which is consistent with the result in Ref.~\cite{Jezo:2014kla}. 

For lower mediator masses, the effect of resonance becomes more important and may dominate over the CC process similar to the Glashow resonance at $E^{}_{\nu} = 6.3~{\rm PeV}$. However, on the one hand, both the flux and sensitivity of tau neutrino telescopes drops at lower neutrino energies. 
On the other hand,  
the limits of non-standard neutrino interactions (NSIs) will come into play if there is a large exotic coupling between electron and neutrino~\cite{Babu:2019vff}.
These limits include the LEP experiment, BOREXINO, IceCube, etc.
Among them, the strongest one is found to be from IceCube atmospheric neutrino data~\cite{IceCubeCollaboration:2021euf}, which set $y/M^{}_{H'} < 0.2\, (100~{\rm GeV})^{-1}$~\cite{Babu:2019vff}.
For a lower incoming neutrino energy, say $E^{}_{\nu} = 0.1~{\rm EeV}$, the resonance is at $M^{}_{H'} = \sqrt{s} \approx  319~{\rm GeV}$ with $y<0.6$, the cross section averaged over a decade in energy turns out to be $8 \times 10^{-34}~{\rm cm}^2$, negligible compared to the SM CC one $\sigma^{}_{\rm CC}(0.1~{\rm EeV}) \approx 4\times 10^{-33}~{\rm cm}^2$.

In the meantime, the $t$-channel exchange of a vector mediator appreciates the forward divergence.
Taking the charged mediators $H'^-$ and $W'^-$ for example, the cross sections for $\nu^{}_{\tau} + e^- \to  e^- + \nu^{}_{\beta}$ are found to be 
\begin{eqnarray}
\sigma^{}_{H'}(s) & = &\frac{Y^{2}_{\tau e} Y^{2}_{\beta e} \left[\frac{s \left(2 M^2_{H'}+s\right)}{M^2_{H'} +s} + 2 M^2_{H'} \ln\frac{M^2_{H'}  }{M^2_{H'} +s}\right]}{32 \pi  s^2} \;, \\ 
\sigma^{}_{W'}(s) & = & \frac{g^{2}_{\tau e} g^{2}_{\beta e} \left[2 \left(M^2_{W'}+s\right)^2 \ln \left(\frac{M^2_{W'}}{M^2_{W'}+s}\right)+s \left(2\frac{ M^4_{W'}+s^2}{M^2_{W'}}+3 s\right)\right]}{8 \pi  s^2 \left(M^2_{W'}+s\right)}\;.
\end{eqnarray}
For small mediator masses, the vector case ($W'^-$) features a forward enhancement, i.e., 
$\sigma \sim g^{2}_{\tau e} g^{2}_{\beta e}/(4\pi M^2_{W'})$ 
with $M^{}_{W'} \ll \sqrt{s}$. This, however, does not apply to the spin-flipping scalar mediator ($H'^-$).
The NSI limits will also apply to the $t$-channel processes.
Recalling $y/M < 0.2\, (100~{\rm GeV})^{-1}$, the optimal cross sections can be estimated by taking $g^{}_{\tau e} =  g^{}_{\beta e} =1$ and $M^{}_{H'} = M^{}_{W'} = 500~{\rm GeV}$, which yields $\sigma^{}_{W'} \sim 8 \times 10^{-35}~{\rm cm^2}$ and $\sigma^{}_{H'} \sim 2 \times 10^{-36}~{\rm cm^2}$ for $E^{}_{\nu} = 1~{\rm EeV}$.
Similar conclusions can be made for neutral mediators, e.g., $Z'$. But, in this case the forward enhancement of vector force (i.e., small momentum transfer) does not noticeably alter the final-state neutrino energy, which weakens the effect significantly.

Hence, we conclude that both $s$- and $t$-channel neutrino-electron scatterings do not play significant roles at tau neutrino telescopes.
We give in the right panel of Fig.~\ref{fig:xsec} cross sections of $s$- and $t$-channel processes with optimal mediator choice saturating the experimental limits. This visually demonstrates the negligible effect compared to hadronic processes, which are irreducible backgrounds for the neutrino-electron scattering.

\section{Numerical Results} \label{sec:V}
After implementing the neutrino and tau propagation algorithm into GRAND, POEMMA and Trinity setups, we are ready to investigate the modified neutrino interaction induced by new physics effects.
For the GRAND experiment, we assume the matter in the mountain and Earth is composed of uniform standard rock with density $\rho = 2.65~{\rm g \cdot cm^{-3}}$. This is a good approximation for GRAND, as the largest depth in Earth reachable by Earth-skimming neutrinos is around $24~{\rm km}$ with an elevation angle of $5^{\circ}$. For POEMMA, we need to solve the propagation equations with a realistic Earth matter profile, for which we will adopt the PREM model.

\subsection{New physics effects}
Because of the decay and energy loss of taus, the largest length scale  $L^{}_{\tau}$ that a tau can travel in  medium is around $50~{\rm km}$ for all energies. Thus, the tau events registered in the telescope are essentially produced within a layer of thickness $50~{\rm km}$  close to the matter surface. Beyond this thickness, taus converted from $\nu^{}_{\tau}$ can only play a role in the regeneration of $\nu^{}_{\tau}$ via tau decays.
In contrast, neutrinos  at EeV energies possess a much longer mean free path of $L^{\rm CC}_{\nu}\approx 560~{\rm km} \cdot (E^{}_{\nu}/{\rm EeV})^{-0.363} \gg L^{}_{\tau} \approx 50~{\rm km}$ in the standard rock.
Hence the final event number is basically dependent on the nearly-constant neutrino flux strength within $L^{}_{\tau}$ underneath the matter.

The neutrino event number can be roughly estimated with a relation $ N \approx \Phi^{\rm out}_{\nu} \cdot \sigma^{\rm CC}_{\tau}\cdot L^{}_{\tau}$, where $\Phi^{\rm out}_{\nu} $ is the neutrino flux near the surface, $\sigma^{\rm CC}_{\tau}$ is the tau production cross section and $L^{}_{\tau}$ is the tau existing length scale.
Considering two well-separated length scales of taus and neutrinos, the presence of new physics can alter the tau neutrino event number through two possible effects, as has been previously noted. 
First, the new physics enhancement of CC cross section will increase the production rate of taus in the layer close to the surface, e.g., by $\sim \delta \sigma^{\rm CC}_{\tau} \cdot L^{}_{\tau}$ with $\delta\sigma^{\rm CC}_{\tau}$ denoting the possible deviation induced by new physics.
Second, the neutrino flux will experience attenuation before reaching the matter layer near the Earth surface, by a factor $\Phi^{\rm out}_{\nu} \propto \mathrm{exp}\left(-\sigma^{}_{\rm tot} \cdot L^{}_{\nu} \right)$, where $L^{}_{\nu}$ is the distance traveled by neutrinos in matter and $\sigma^{}_{\rm tot}$ is the total cross which depletes the neutrino flux.
Note that $\sigma^{}_{\rm tot}$ contains both the CC interacting producing taus as well as processes not related to the tau production, e.g., the NC interaction.
The modification in the final event number due to new physics can be estimated by
\begin{eqnarray} \label{eq:deltaN}
\delta N & = & \Phi^{\rm in}_{\nu} \cdot L^{}_{\tau}  \mathrm{e}^{-\sigma^{}_{\rm tot}  \cdot L^{}_{\nu}}  \cdot\left( \delta \sigma^{\rm CC}_{\tau} - \sigma^{\rm CC}_{\tau}\, L^{}_{\nu}\, \delta \sigma^{\rm }_{\rm tot}\right) ,
\end{eqnarray}
where $\Phi^{\rm in}_{\nu}$ is the initial neutrino flux and $\delta \sigma^{\rm }_{\rm tot}$ is the modification of total neutrino cross section due to new physics.
Whether we have an increased or decreased event number depends on the sign of $\delta \sigma^{\rm CC}_{\tau} - \sigma^{\rm CC}_{\tau}\, L^{}_{\nu}\, \delta \sigma^{\rm }_{\rm tot}$, which can be simplified to $1- \sigma^{\rm CC}_{\tau} L^{}_{\nu} $
in the assumption of $\sigma^{\rm }_{\rm tot} =  \sigma^{\rm CC}_{\tau}$.
In practice, the telescope will monitor a wide field of view corresponding to a range of neutrino baselines $L^{}_{\nu}$ in matter. We can therefore resolve the presence of new physics by exploring the variance of events as a function of $L^{}_{\nu}$.

\begin{figure}[p!]
	\begin{center}
	\includegraphics[width=0.39\textwidth]{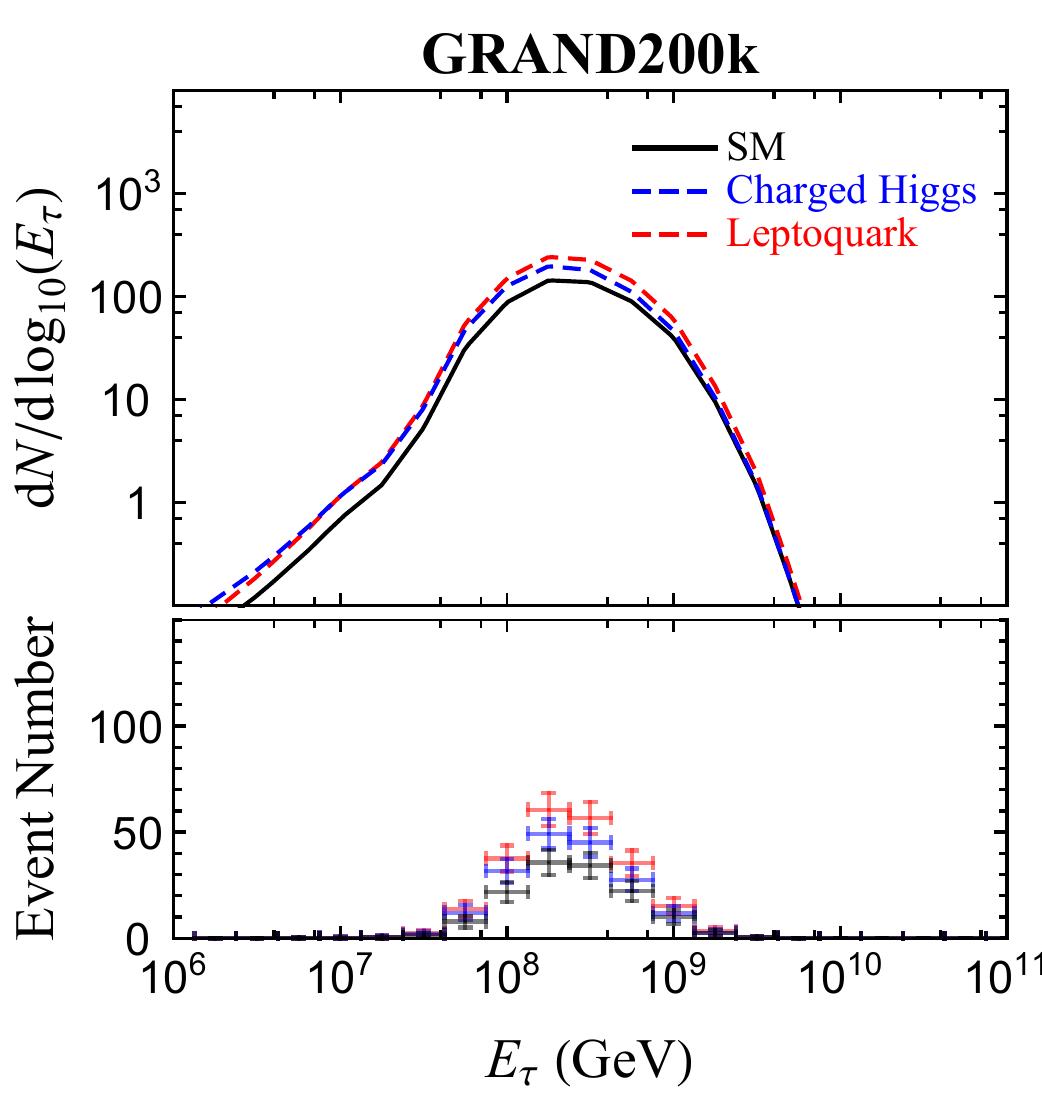}
	\hspace{0.5cm}
	\includegraphics[width=0.39\textwidth]{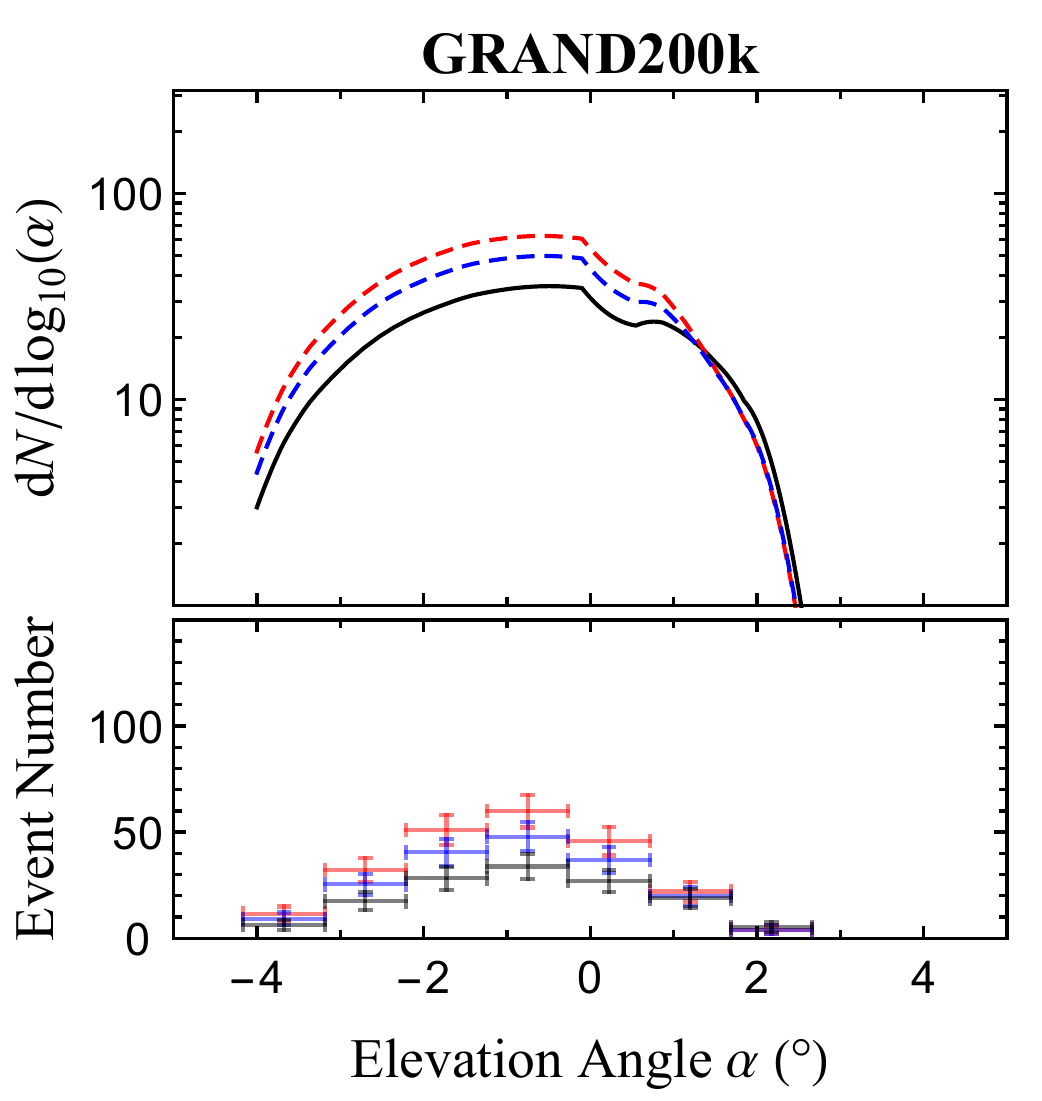}
		\includegraphics[width=0.39\textwidth]{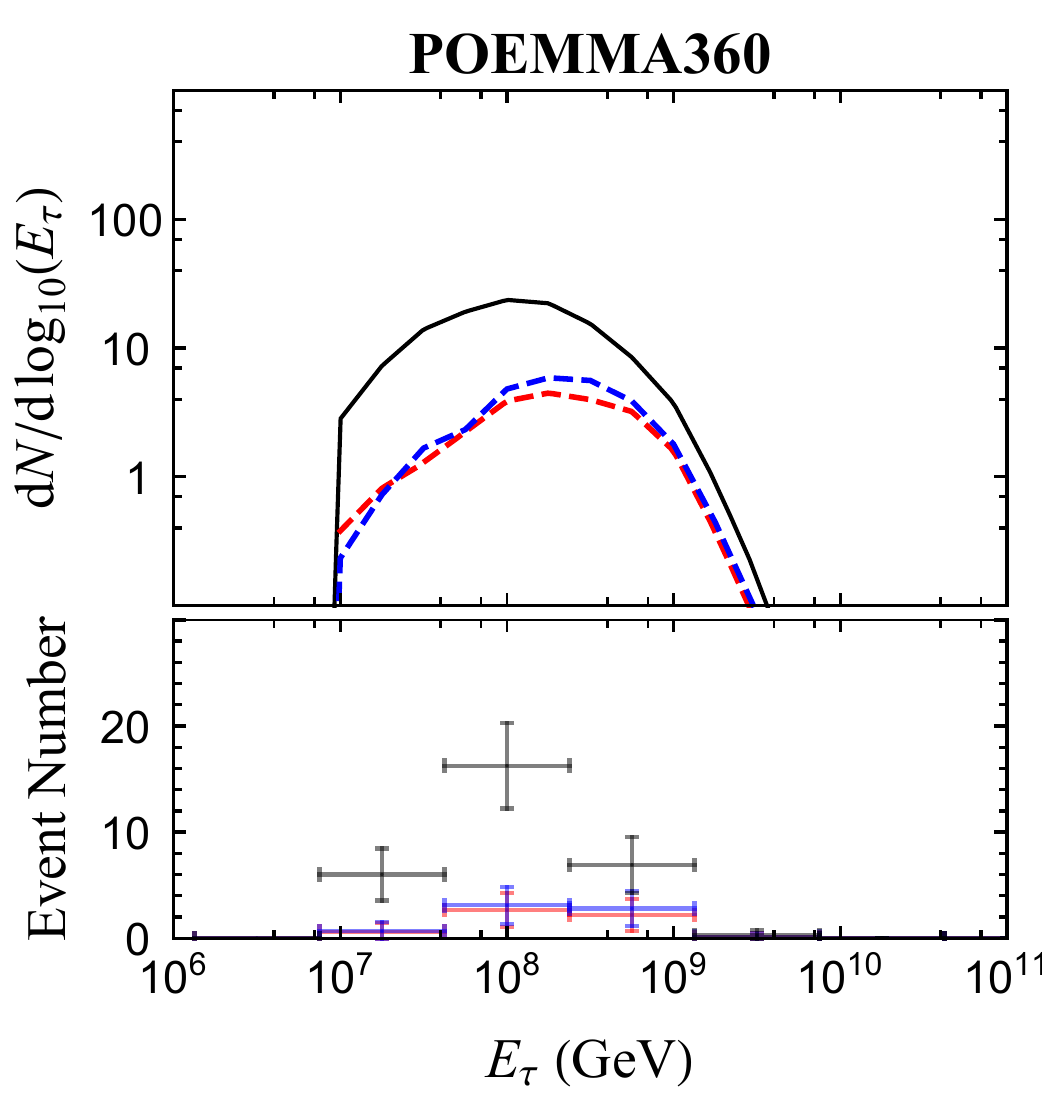}
		\hspace{0.5cm}
		\includegraphics[width=0.39\textwidth]{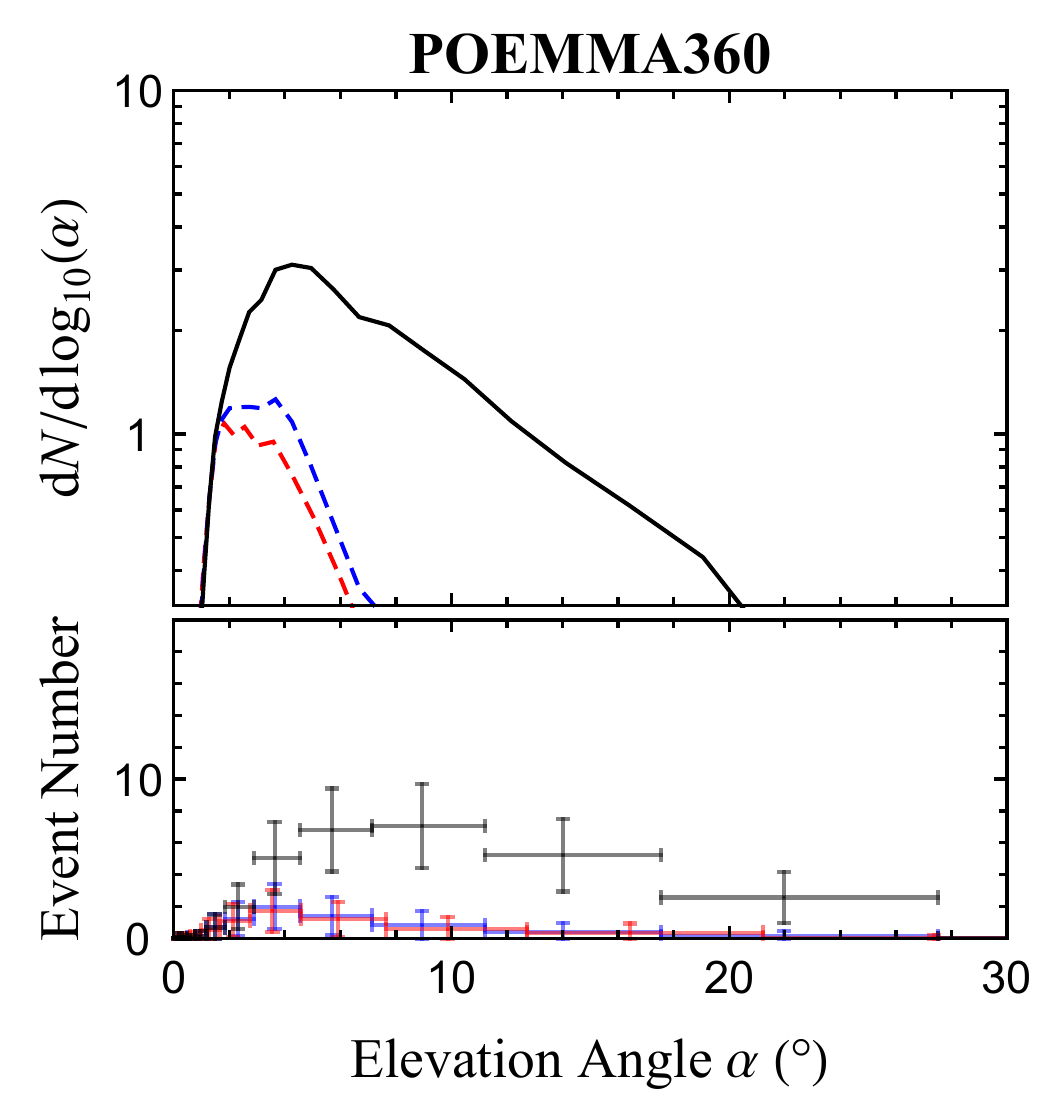}

		\includegraphics[width=0.39\textwidth]{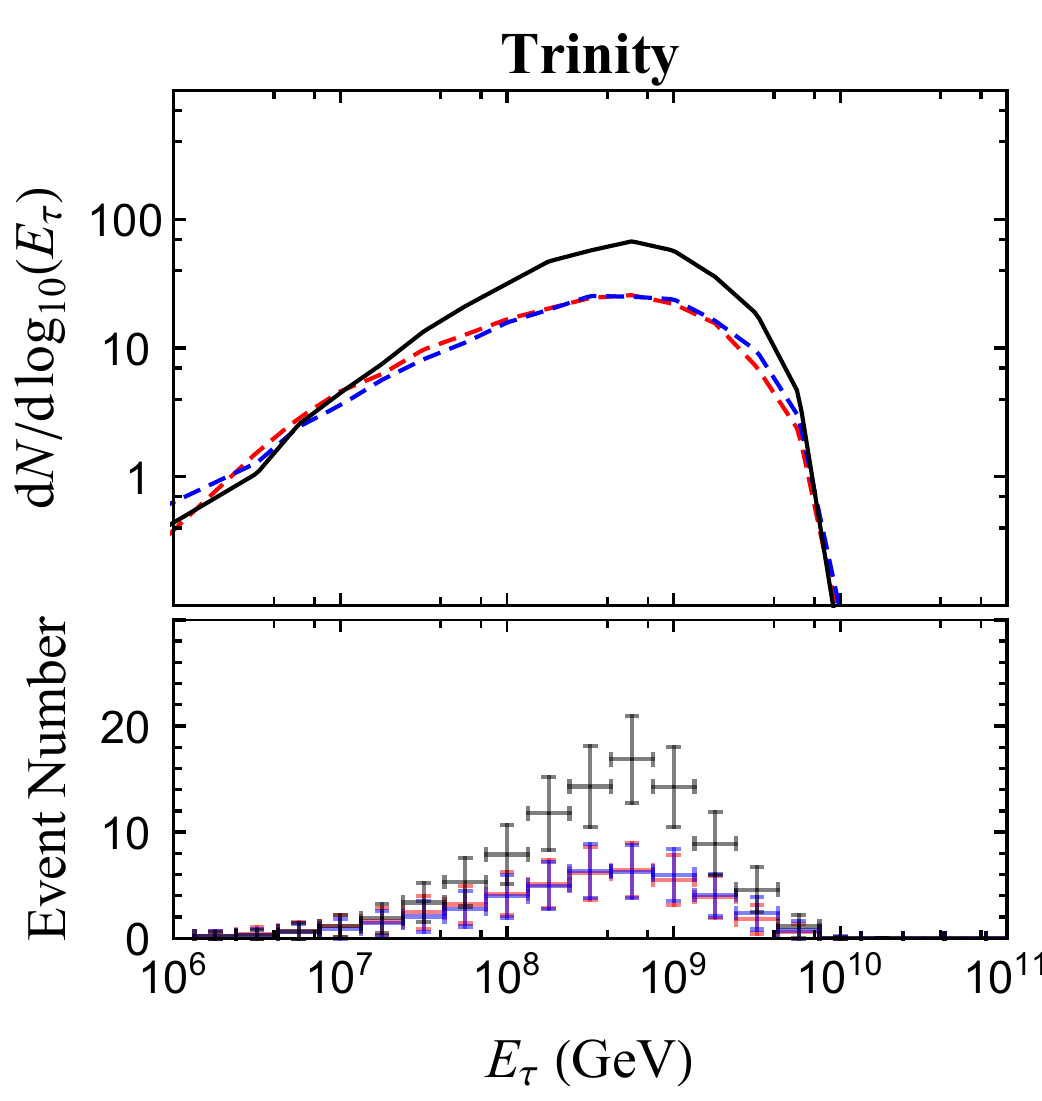}
		\hspace{0.5cm}
		\includegraphics[width=0.39\textwidth]{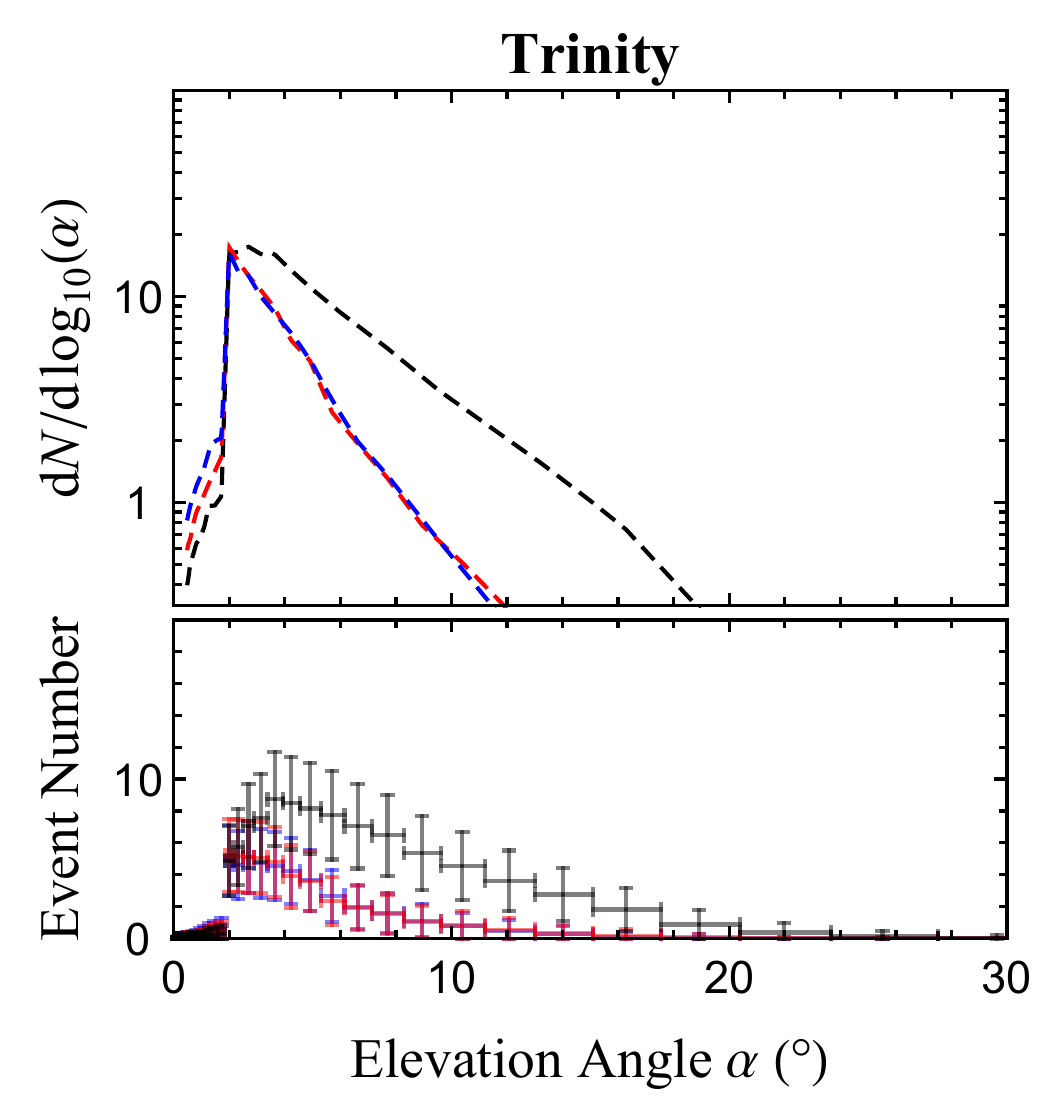}
	\end{center}
	\vspace{-0.3cm}
	\caption{The energy (left panels) and angular (right panels) distributions of tau neutrino events at GRAND, POEMMA and Trinity experiments. The black curves represent the SM case, while the blue and red ones stand for the scenarios of charged Higgs ($M^{}_{ H'} = 90~{\rm GeV}$ and $y^{q}_{\rm cs} = y^{\ell}_{\tau\tau} = 1$)  and leptoquark ($M^{}_{\rm LQ} = 400~{\rm GeV}$ and $y^{\rm LL}_{{\rm s} \tau} = y^{\rm RR}_{{\rm c} \tau} = 1$ for $S^{}_{1}$ leptoquark), respectively. The initial diffuse neutrino flux is fixed as in Ref.~\cite{Murase:2015ndr} for demonstration. }
	\label{fig:distribution}
\end{figure}
The energy and angular distributions of event number in GRAND, POEMMA and Trinity for two new physics scenarios are illustrated in Fig.~\ref{fig:distribution}.
In the top two panels, we give the event distributions for GRAND200k assuming the elevation angle of antenna screen is $3^{\circ}$.
The middle and bottom panels show the distributions of POEMMA360 and Trinity (three stations), respectively, and both of them have the full FOV in azimuth. 
In all panels, the black curves stand for the SM scenario, while the blue and red ones are for the charged Higgs with $M^{}_{H'} = 90~{\rm GeV}$ and $y^{}_{\rm cs}= y^{}_{\tau\tau} =1 $ and leptoquark with $M^{}_{\rm S_{1}}= 400~{\rm GeV}$ and $y^{\rm LL}_{1,{\rm s\tau(c \tau)}} = y^{\rm RR}_{1,{\rm c \tau}} = 1$, respectively.
In each panel, both the differential distribution (upper one) with respect to the tau energy or the elevation angle and the binned event number with error bars (lower one) are given. 
In the following, we make some observations on Fig.~\ref{fig:distribution}.
\begin{itemize}
	\item With the given setup, GRAND200k can collect much more events than POEMMA360, as one can expect from their sensitivities to diffuse neutrino flux in Fig.~\ref{fig:fluxSens}. 
	For the angular distribution in the right panel, the elevation angle $\alpha < 0^{\circ} $ corresponds to the event coming from above the horizon, which transverses only the mountain target. 
	Because the typical mountain thickness $\sim 100~{\rm km}$ is much smaller compared to the neutrino attenuation length, events with $\alpha < 0^{\circ} $ are very useful to normalize the initial neutrino flux.
	On the other hand, an event with $\alpha > 0^{\circ} $ (Earth-skimming neutrinos) corresponds typically to a longer traveling distance. 
	The attenuation effect is what we will use to extract the information of the neutrino cross section. 
	On average, the additional contributions from leptoquark and charged Higgs models will increase the event rate at GRAND, and distort the angular distribution of events. Compared to GRAND, Trinity can collect more Earth-skimming neutrinos.
	\item The angular distribution of POEMMA events reaches the maximum around the elevation angle $\alpha = 4^{\circ}$, and then it drops rapidly as we go to larger elevation angles due to the attenuation effect.
	Note that for POEMMA there is a one-to-one correlation between the elevation angle and the distance traveled by neutrinos in Earth.
	In comparison to GRAND,
	the new physics contribution reduces the event number significantly at POEMMA. This is due to that the majority of FOV from the POEMMA satellite corresponds to very long chord lengths in the Earth. The event modification due to new physics  is negative in Eq.~(\ref{eq:deltaN}) with $L^{}_{\nu} \gg 1 / \sigma^{\rm CC}_{\tau} $.
	\item The energy distribution of events at a telescope is determined by two factors: (i) the input of initial neutrino flux, which has been taken from Ref.~\cite{Murase:2015ndr}; (ii) the average distance traveled by neutrinos. 
	Because neutrinos interact more strongly at higher energies, one can observe a shift of event to lower energies as the neutrino flux has experienced multiple scatterings over a very long distance. This effect can be reflected by the comparison between  energy distributions of GRAND and POEMMA, the latter of which has a longer average baseline.
\end{itemize}
It should be stressed that the GRAND, POEMMA and Trinity experiments are remarkably complementary, because the new physics enhances the event rate at one experiment but reduce the rate at another. Their combination will be very useful to maximize the sensitivity to new physics. Next, we shall investigate the new physics effect in a statistically quantitative approach.

\begin{figure}[t!]
	\begin{center} 
		\includegraphics[width=0.4\textwidth]{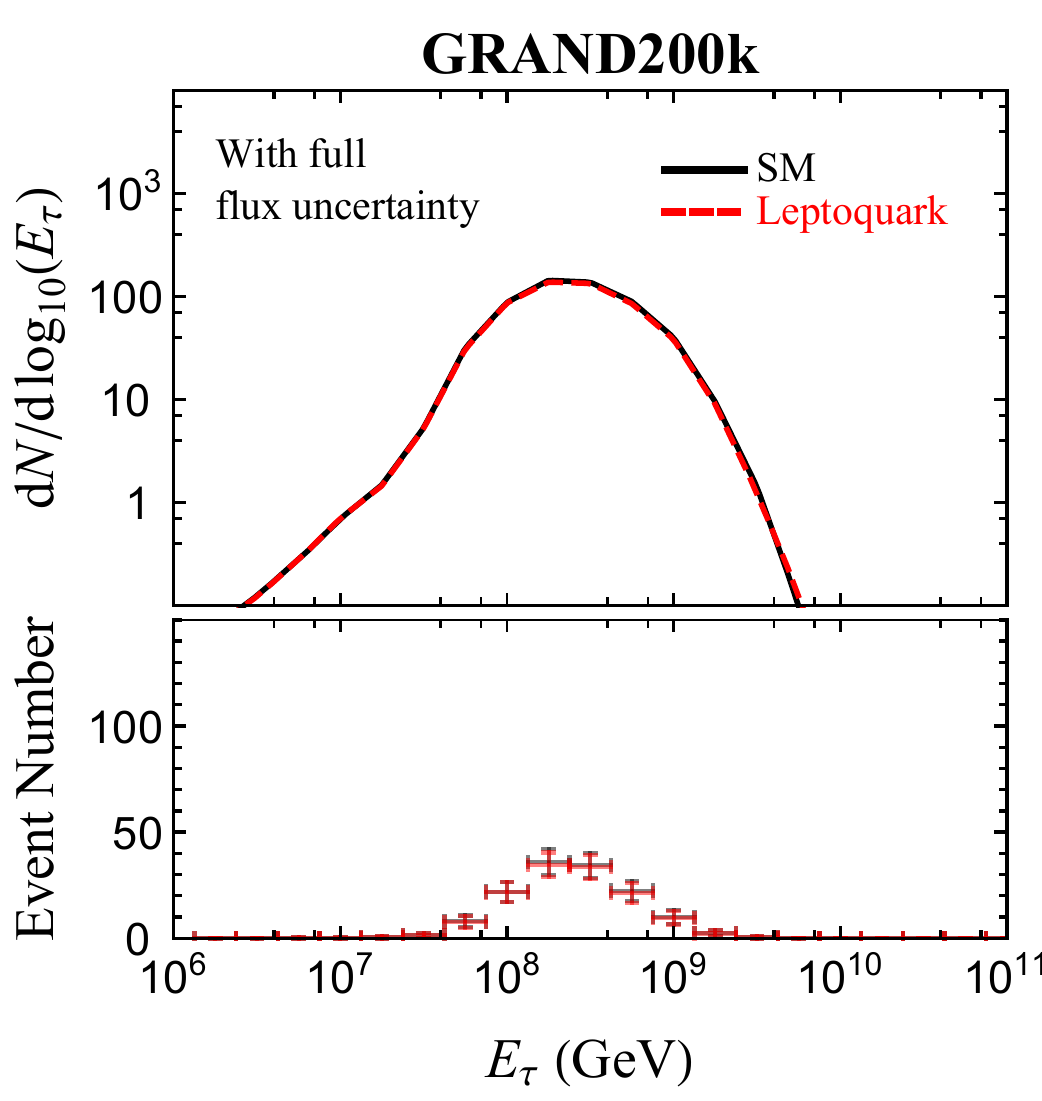}
		\hspace{0.5cm}
		\includegraphics[width=0.4\textwidth]{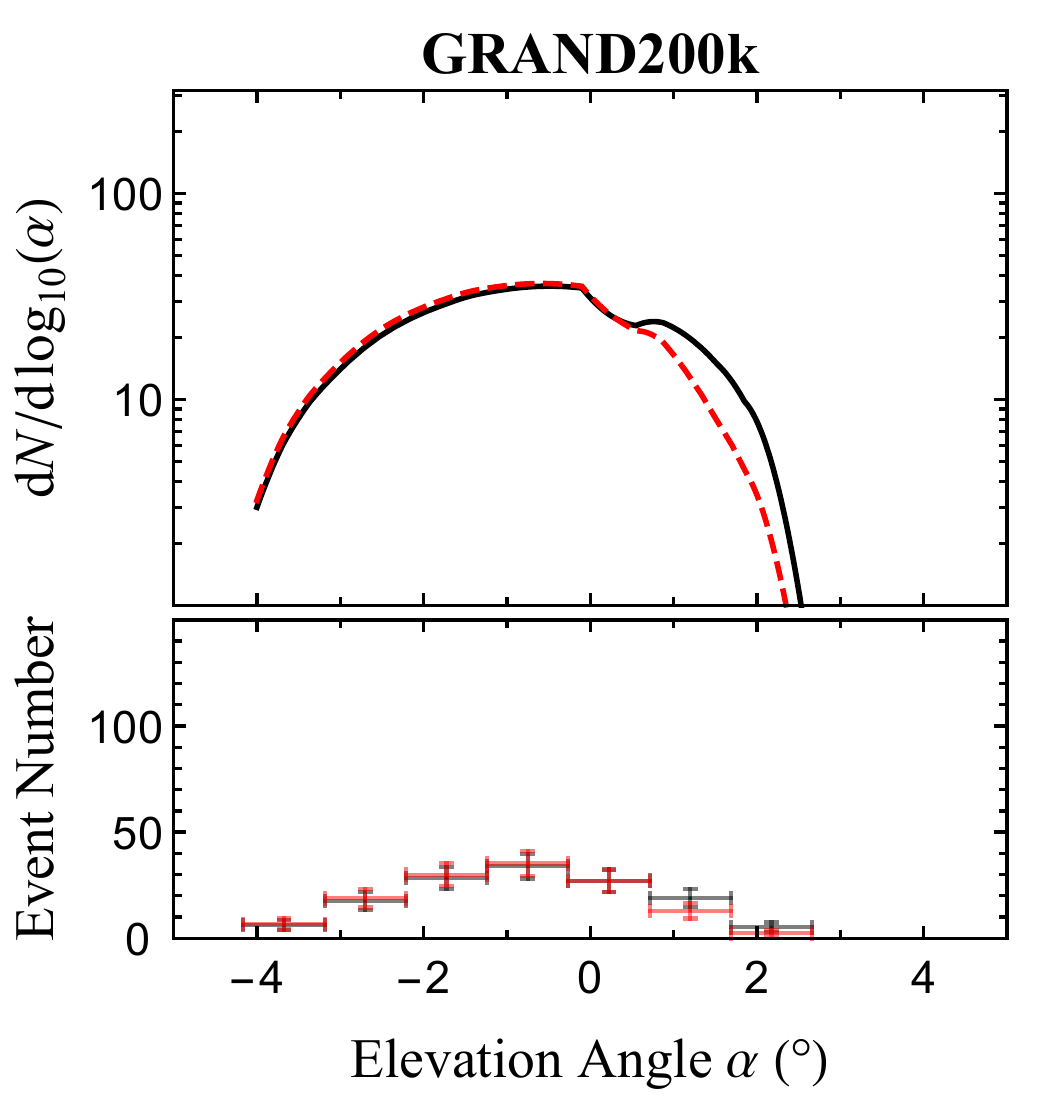}
	\end{center}
	\vspace{-0.3cm}
	\caption{The energy (left panel) and angular (right panel) distributions of the SM and the leptoquark cases. In comparison to Fig.~\ref{fig:distribution}, their differences are minimized by marginalizing over the unknown initial flux of cosmogenic neutrinos.}
	\label{fig:initialFlux}
\end{figure}

\subsection{Sensitivities}
The largest systematic uncertainty stems from the unknown priors of the diffuse neutrino flux.
The magnitude and shape of the cosmogenic neutrino flux is partly model-dependent, resting on, e.g., the evolution model of cosmic sources and the initial chemical component of cosmic rays.
To be conservative, we can assume that the initial tau neutrino flux is completely unknown, which is to be derived from the same data set at tau neutrino telescopes that we use to probe the new physics models. 
This induces a degeneracy between the new physics effect and the initial flux input. For instance, a larger or smaller input of initial neutrino flux can mimic the effect of event excess or depletion due to new physics.
The key to resolve the degeneracy is relying on the angular distribution of neutrino events at tau neutrino telescopes.
In Fig.~\ref{fig:initialFlux}, we show as an example the event distributions at GRAND by varying the initial neutrino flux such that the deviation between the SM and new physics is minimized.
We find that the new physics effect in energy distributions can be completely compensated by varying the input of the unknown initial neutrino flux. In comparison, the difference in angular distributions is stable against the change in the initial flux.
In order to quantify the flux uncertainty, we split the initial neutrino flux into 24 bins in the energy range of $(10^6, 10^{12})~{\rm GeV}$. 
Four bins form a decade in energy, which should be fine enough to control the variation in the cosmogenic flux 
\footnote{Even if the variation in cosmogenic neutrino flux could be finer than the bin size we adopt, the result does not make any noticeable difference, because neutrinos within each bin almost have the identical transport behavior and similar amount of energy deposition in the detector.}.
We then let the magnitude of neutrino flux in each energy bin vary freely while fitting the simulated data.

The further observation of cosmic rays will no doubt enhance our knowledge of the nature of cosmic rays, and hence reduce the systematical uncertainty of neutrino flux. 
As another benchmark, we shall also explore the sensitivity by assuming the initial neutrino flux to be completely fixed by cosmic ray observations.
The true sensitivity in the future experiment should be in between this ideal case and the conservative treatment above.

Another systematic error originates in the PDF uncertainties. 
This can be conveniently evaluated by using the available PDF set with errors~\cite{Clark:2016jgm}.
To accommodate this uncertainty, we repeat the whole computation for all the 59 PDF sets from the CT18 PDFs~\cite{Hou:2019efy}, and take the difference in each computation as the error induced by the corresponding PDF. The uncertainty of SM CC cross section in Fig.~\ref{fig:xsec} is calculated in this way, and a similar procedure will be performed for the final events.

\begin{figure}[t!]
	\begin{center}
		\includegraphics[width=0.43\textwidth]{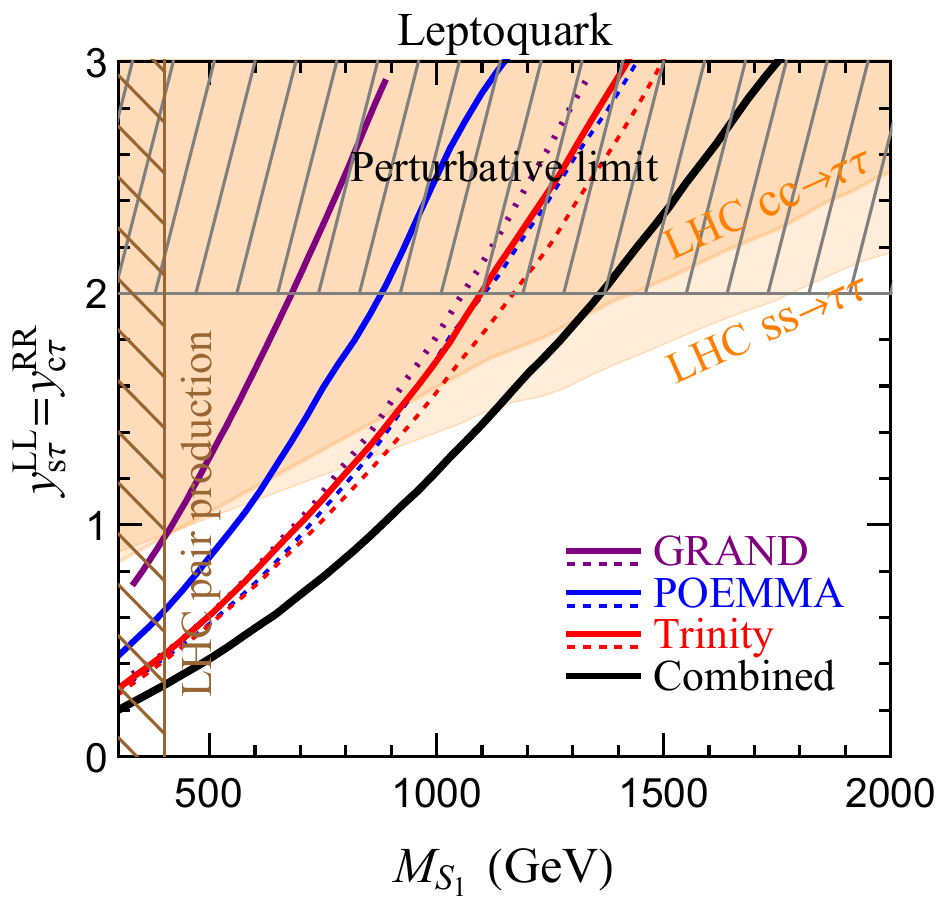}
		\includegraphics[width=0.43\textwidth]{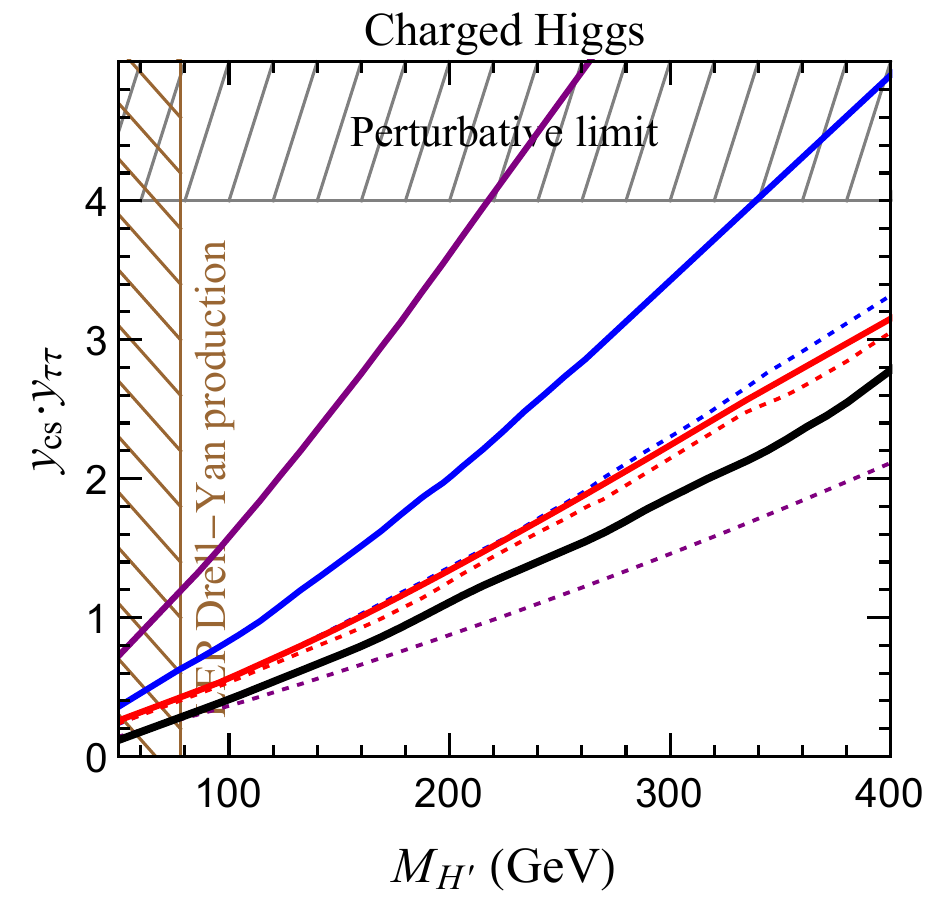}
	\end{center}
	\vspace{-0.3cm}
	\caption{The sensitivities of GRAND (purple curves), POEMMA (blue curves) and Trinity (red curves) to the leptoquark (left panel) and charged Higgs (right panel) models, at $90\%$ confidence level. The black curves stand for the combined sensitivity of three telescopes. The dotted curves give the results without considering the uncertainty of initial neutrino flux.
	For comparison, we also give the perturbative limit of Yukawa couplings, the  LEP~\cite{ALEPH:2013htx} and LHC~\cite{Angelescu:2018tyl} constraints on the lower masses of charged Higgs and leptoquark, as well as the leptoquark constraint from LHC high-$p^{}_{\rm T}$ dilepton tail searches.
}
	\label{fig:sens}
\end{figure}

The energy resolution of POEMMA is subject to the long distance from the emerging point of tau to the satellite camera.
Owing to the rapid scattering of photons off atmospheric molecules and aerosols, POEMMA has a very limited energy resolution to the extensive air shower induced by tau decay.
The  energy resolution of tau for $\lesssim 1~{\rm EeV}$ is not provided by the POEMMA collaboration, but we can estimate it according to the empirical relation $\Delta E/ E \approx 1/\sqrt{N^{}_{\rm PE}} \lesssim 30\%$~\cite{Neronov:2016zou}. The fluctuation of air shower development might worsen the energy reconstruction. A conservative choice of the resolution $\Delta E/ E = 100\%$ will be adopted in this work. 
The angular resolution of POEMMA is limited by the Cherenkov cone, typically $1.5^{\circ}$ in the view of the satellite camera.
This can be translated into the resolution on the tau elevation angle according to the relation
$\alpha = \arcsin{\left[(R^{}_{\oplus}+ h^{}_{\rm s})/R^{}_{\oplus} \sin{\theta^{}_{\rm s}} \right]} - 90^{\circ}$, with $\theta^{}_{\rm s}$ being the viewing angle of the satellite with respect to the vertical.
For $\Delta \theta^{}_{\rm s} = 1.5^{\circ}$ near the horizon, the difference in tau elevation angle is as large as $\Delta \alpha \sim 8^{\circ}$, due to the very high altitude of POEMMA satellite.
The ground-based GRAND and Trinity experiments outperform POEMMA in both energy and angular resolutions. For instance, the energy resolution of GRAND can be as good as $15\%$, and the angular reconstruction can achieve a level of sub-degree (less than $0.5^{\circ}$) on average~\cite{GRAND:2018iaj}.

By summing over the two-dimensional grid of energy and elevation angle, we obtain the minimum of $\chi^2$ via
\begin{eqnarray}
\chi^2_{\rm min}= \underset{\Phi^{\rm in}_{\nu}}{\rm Min} \left\{ \sum_{i = 1}^{N_{\rm bins}} \frac{\left( n_{i}^{\rm th} - n_{i}^{\rm exp} \right)^2}{n_{i}^{\rm th} + \sigma^{2}_{{\rm PDF},i}} \right\},
\end{eqnarray}
where $n^{\rm exp}_{i}$ and $n^{\rm th}_{i}$ are the nominal experimental event number and the theoretical prediction from new physics models, respectively, in the $i$-th bin. The total bin number in the energy and angle grid is $N^{}_{\rm bins}$.
Note that for our current sensitivity analysis, we do not include higher order processes which in our analysis will appear as part of the  theoretical systematics~\cite{Soto:2021vdc}.
To generate the experimental data $n^{\rm exp}_{i}$, we use the cosmogenic neutrino flux in Ref.~\cite{Murase:2015ndr} as the input, and adopt the SM cross section without new physics, i.e., assuming the SM as the true model.
For each given parameter choice of new physics, the theoretical expectation $n_{i}^{\rm th}$ can be calculated with a randomly given diffuse neutrino flux.
The minimum of chisquare, $\chi^2_{\rm min}$, will be obtained by scanning over initial neutrino fluxes, which is expected to reduce the statistical significance due to the degeneracy between flux and new physics effects.
The value of $\chi^2_{\rm min}$ can then be used to constrain new physics models regardless of priors of the diffuse neutrino flux.

In Fig.~\ref{fig:sens}, the purple, blue and red solid curves represent the sensitivities of GRAND, POEMMA and Trinity, respectively, to the charged Higgs (left panel) and leptoquark (right panel) models. 
The combined sensitivity of those three experiments is given as the black curves.
The existing laboratory constraints on the scenario we are considering are also presented for comparison.
We observe that these three telescopes can have sensitivities surpassing the current collider limits.
Some further comments on the results are given below.
\begin{itemize}
	\item The mountain-based telescopes GRAND and Trinity have comparable sensitivities to the charged Higgs and leptoquark models, if the initial diffuse neutrino flux is given as a well-known theoretical prior. 
	However, as for the case with unknown initial flux, Trinity has the best sensitivity.
	Though the effective exposure of GRAND is the largest among three telescopes, 
	many events of GRAND are coming from the direction above the horizon (penetrating only the mountain), where the degeneracy between the new physics contribution and flux uncertainty is difficult to resolve. 
	Whereas, Trinity collects exclusively the Earth-skimming neutrinos which contain more information about neutrino interactions. Since the final sites of GRAND are yet to be determined, our result here only represents a special scenario with the antenna screen being deployed on a $3^{\circ}$ inclined mountain. A steeper host slope  for GRAND, which has a wider FOV for Earth-skimming neutrinos, will certainly improve the results here.
	\item The POEMMA experiment is subject to the high altitude of  satellite, i.e., $525~{\rm km}$, in comparison to the mountain-based telescope $\sim 2~{\rm km}$. A higher altitude will worsen the  resolution of neutrino elevation angle, while the new physics mostly manifest itself by altering the angular distribution of events. 
	Therefore, the ground-based Trinity setup seems to be more optimized in probing the neutrino interaction of our concern.
\end{itemize}
The combination of GRAND, POEMMA and Trinity greatly enhances the sensitivity. Instead of a simple sum of $\chi^2_{\rm min}$, these three telescopes are complementary to each other in resolving the diffuse flux uncertainty and probing the new physics effect.
The above discussions are made for the case with the initial neutrino flux unknown.
As a comparison, we also give the ideal case where the initial neutrino flux is completely known and not minimized over as the dotted curves in Fig.~\ref{fig:sens}.

\section{Conclusions}\label{sec:VI}

Other than supplementing the multimessenger astronomy, tau neutrino telescopes can also play the role of a particle collider which collides a high-energy neutrino beam with proton and electron.
We have systematically investigated the new physics scenarios that modify the neutrino-matter interactions relevant for tau neutrino telescopes including: charged and neutral Higgs, leptoquark, as well as neutral and charged gauge bosons. These extended scenarios have already been under tight constraints from ground-based colliders like LEP and LHC. Among them, we find the charged/neutral Higgs and leptoquark can have significant imprint on the neutrino-proton scattering, if the existing experimental limits are considered. 
In particular, this will require the new particles to exclusively have large couplings with the second or third family, for which the related processes can be suppressed at LEP or LHC. Tau neutrino telescopes probe the neutrino-proton COM energy as high as 45~TeV, where the PDFs of second and third generations of quarks are not that suppressed and in some case can be comparable to that of u and d.
The gauge bosons are mainly subject to feasible model construction which constrains the lower mass together with collider searches.

By solving the neutrino and tau propagation equations, we have generated the events at GRAND, POEMMA and Trinity with configurations close to their realistic experimental setups. Their sensitivities to the new physics scenarios are limited by the unknown prior of the cosmogenic neutrino flux, which might be improved with future observations of cosmic rays. Under the assumption of completely unknown flux priors, we find that the angular distribution of events carries most information about the absolute cross section of neutrinos. With a $\chi^2$-analysis, we have obtained the sensitivities of three representative tau neutrino telescopes to the parameter space of charged Higgs and leptoquark models, i.e., Fig.~\ref{fig:sens}.
We should keep in mind that those sensitivity curves are expected to shift according to the final design of these three telescopes.

Tau neutrino telescopes, as a particle collider, feature highest neutrino-matter colliding energies. However, we need to point out that their sensitivity to the new physics scenario (in particular, those which have been explored in the work) is more or less restricted for a number of reasons. First, the initial neutrino flux is not under control and spreads over a wide range, compared to the nearly monoenergetic collider beam, which will partly erase the potential new physics signatures.
Second, even though we also have an electron target in matter, neutrino-proton collision will produce a large irreducible background. This restriction limits the new physics searches through the clean pure leptonic portal.
Third, the event topology at tau neutrino telescopes is not as broad as in the collider spectrometer, and the detailed product of neutrino-matter collision is not a direct observable.

Nevertheless, we want to point out that tau neutrino telescopes are very sensitive to certain new physics scenarios. For instance, if the neutrino collision product is a long-lived exotic particle, the signal similar to the anomalous events at ANITA can be induced.
Another interesting possibility is the presence of double or multiple cascade events with $(2+n)\tau$ final states at tau neutrino telescopes, which can appear in certain processes. This special topology is easy to resolve  with negligible SM leading-order background, given good enough time resolution at the detection array.
It is also promising to probe other processes with SM unknowns or new physics beyond at tau neutrino telescopes including: 
sphalerons~\cite{Ellis:2016dgb}, 
QCD saturation effect~\cite{Golec-Biernat:1998zce,Bartels:2002cj,Iancu:2003ge,Armesto:2007tg,Illarionov:2011wc,Block:2014kza,Arguelles:2015wba}, 
test of equivalence principle and Lorentz invariance~\cite{Coleman:1997xq,Gonzalez-Garcia:2005ryx,Arguelles:2015dca,Gonzalez-Garcia:2006koj,Mattingly:2009jf,Murase:2009ah,IceCube:2010fyu,Gorham:2012qs,Esmaili:2014ota,Wang:2016lne,Arguelles:2016rkg,Liao:2017yuy,Stecker:2017gdy,IceCube:2017qyp,Zhang:2018otj,Fiorillo:2020gsb,Chianese:2021vkf,Arguelles:2021kjg,IceCube:2021tdn},
microscopic black holes~\cite{Uehara:2001yk,Alvarez-Muniz:2002snq,Dutta:2002ca,Kowalski:2002gb,Jain:2002kz,Stojkovic:2005fx,Illana:2005pu,Anchordoqui:2006fn,Kisselev:2010zz,Arsene:2013nca,Reynoso:2013jya,Mack:2019bps}, 
neutrino transition magnetic moment~\cite{Coloma:2017ppo,Coloma:2019qqj},~etc.

\section*{Acknowledgments}
Authors would like to thank Carlos Argüelles, Aart Heijboer, Matthew Kirk, Steven Prohira, Makoto Sasaki, Stephanie Wissel, Pavel Zhelnin and Shun Zhou for useful comments and communications.
The Feynman diagrams are generated using  Jaxodraw~\cite{Binosi:2003yf}.
GYH is supported by the Alexander von Humboldt Foundation.

\appendix

\section{Neutrino Cross Section: Standard Model}
The expressions of neutrino cross sections in the SM are partly available in the literature~\cite{Gandhi:1995tf,Gandhi:1998ri,Formaggio:2012cpf,NuSigma,Bertone:2018dse}, but we present them here for completeness.
For a general process $\nu + {\rm N} \to l + q^{}_{\rm f}$, 
the ultimate differential cross section with respect to the Bjorken scaling variables
can be obtained by converting from the parton-level cross section in the center-of-mass frame.
In the following expressions, we assume the most general cases which do not require the final states $l$ and $q$ to be massless. For $q=$ u, d, s, c and b, it is a good approximation to set $m^{}_{q} = 0$ for our concerned neutrino energy.
But for top quark, the production threshold corresponds to a Bjorken-scaling variable $ m^{2}_{\rm t} /(2 E^{}_{\nu} M) \sim 1.6 \times 10^{-5}$ for $E^{}_{\nu} = 1~{\rm EeV}$, which cannot be neglected at energy scales of our interest.
To incorporate the effect of final-state hadron mass, the usually adopted Bjorken-scaling variable $x$ in PDFs should be replaced by $x'$ in the slow-rescaling prescription\cite{Barnett:1976kh,Gandhi:1995tf,Gandhi:1998ri}:
\begin{eqnarray}
\hat{s} & \equiv & 2 M E^{}_{\nu} x' \;, \\
Q^2 & \equiv & 2 M E^{}_{\nu} x y  = 2 M E^{}_{\nu} x' y  - m^2_{q} \;,
\end{eqnarray}
where $x' = x + m^2_{q} / (2 M E^{}_{\nu} y)$ is the modified Bjorken variable, $y = (1-E^{}_{l}/E^{}_{\nu})$ is the inelasticity, $\hat{s}$ is the square of total energy  in the parton COM frame, and $Q^2 = -t$ is the square of momentum transfer. In the limit of $m^{}_{q} \to 0$, we recover the massless case $x' = x$.

At the parton level, the charged-current cross section between  neutrino and  quark parton in the COM frame reads
\begin{eqnarray}
\frac{\mathrm{d}\sigma^{}_{\nu {D}}}{ \mathrm{d} \cos{\theta}}&  =  & \frac{G^2_{\rm F} M^4_{W} (\hat{s} - m^2_{q_{\rm f}} - m^2_{l})}{2\pi (t - M^{2}_{W})^2} \;,  \\
\frac{\mathrm{d}\sigma^{}_{\nu \overline{U}}}{ \mathrm{d} \cos{\theta}} & =  & \frac{G^2_{\rm F} M^4_{W} (u - m^2_{q_{\rm f}})(u - m^2_{l})}{2\pi (t - M^{2}_{W})^2 \hat{s} } \;,
\end{eqnarray}
for the down- and up-type quarks ${ D}$ and $\overline{U}$, respectively, where $u = \sum_{\rm All} m^{2}_{i} - \hat{s} - t$.
Incorporating PDFs for the quark parton $q(x')$, we have
\begin{eqnarray}
\frac{\mathrm{d^2}\sigma^{}_{\nu {\rm N}}}{  \mathrm{d}x'\; \mathrm{d} \cos{\theta} }  = \sum^{}_{D = {\rm d, s, b}} {D}(x') \frac{\mathrm{d}\sigma^{}_{\nu {D}}}{ \mathrm{d} \cos{\theta}} + \sum^{}_{\overline{U} = {\rm \overline{u},\overline{c}}} \overline{U}(x') \frac{\mathrm{d}\sigma^{}_{\nu \overline{U}}}{ \mathrm{d} \cos{\theta}} \;.
\end{eqnarray}
The inelasticity $y$ is connected to the zenith angle in the parton COM frame with
\begin{eqnarray}
y & = & \frac{\hat{s}  + m^2_{q} - m^2_{l}-\lambda^{1/2}(\sqrt{\hat{s}},m^{}_{l},m^{}_{q}) \cos{\theta}}{2 \hat{s}} \;,
\end{eqnarray}
where $4 \hat{s} |\vec{p}^{}_{l}|^2 = \lambda(\sqrt{\hat{s}},m^{}_{l},m^{}_{q}) =  (\hat{s} - (m^{}_{l} - m^{}_{q})^2)(\hat{s} - (m^{}_{l} + m^{}_{q})^2)$ with $\vec{p}^{}_{l}$ being the momentum of $l$ in the COM frame.
By converting all Mandelstam variables into the Bjorken $x'$ and inelasticity $y$, the final cross section takes the form
\begin{eqnarray}
\frac{\mathrm{d^2}\sigma^{}_{\nu {\rm N}}}{  \mathrm{d} x'\; \mathrm{d} y }  =  \frac{\mathrm{d^2}\sigma^{}_{\nu {\rm N}}}{  \mathrm{d}x'\; \mathrm{d} \cos{\theta} }  \cdot \left|\frac{\partial (x', \cos{\theta})}{ \partial(x',y)} \right| .
\end{eqnarray}
Here, the absolute value of determinant of the Jacobian matrix reads $|{\partial (x', \cos{\theta})}/{ \partial(x',y)}| = |\partial \cos{\theta} /\partial{y}| = 2 \hat{s} / \lambda^{1/2}(\sqrt{\hat{s}},m^{}_{l},m^{}_{q}) $, which yields a standard factor of 2 in the massless limit. The integration limits of $x'$ and $y$ are
\begin{eqnarray}
\frac{(m^{}_{l} + m^{}_{q})^2}{2 M E^{}_{\nu}} & \leq x' \leq  & 1 \;, \\
\frac{\hat{s} +  m^2_{q} - m^2_{l} - \lambda^{1/2}(\sqrt{\hat{s}},m^{}_{l},m^{}_{q})}{ 2 \hat{s}} & \leq y \leq &  \frac{\hat{s} +  m^2_{q} - m^2_{l}  + \lambda^{1/2}(\sqrt{\hat{s}},m^{}_{l},m^{}_{q})}{ 2 \hat{s}} \;.
\end{eqnarray}
One can check the correctness by considering the extreme case $m^{}_{l} = 0$ and $\hat{s} \to m^2_{q}$, such that $q$ is produced nearly at rest in the COM frame. In this case, the hadron $q$ should take away almost all the initial neutrino energy, i.e., $y \to 1$.

The neutral-current cross section at the parton level can be similarly obtained
\begin{eqnarray}
\frac{\mathrm{d}\sigma^{}_{\nu {q}}}{ \mathrm{d} \cos{\theta}}&  =  & \frac{G^2_{\rm F} M^4_{Z} \left[ \left( g^2_{\rm V}+ g^2_{\rm A}\right)^2 \hat{s}^2 + \left( g^2_{\rm V}- g^2_{\rm A}\right)^2 u^2  \right]}{8\pi (t - M^{2}_{Z})^2 \hat{s}} \;,  \\
\frac{\mathrm{d}\sigma^{}_{\nu \overline{q}}}{ \mathrm{d} \cos{\theta}}&  =  & \frac{G^2_{\rm F} M^4_{Z} \left[ \left( g^2_{\rm V}+ g^2_{\rm A}\right)^2 u^2 + \left( g^2_{\rm V}- g^2_{\rm A}\right)^2 \hat{s}^2  \right]}{8\pi (t - M^{2}_{Z})^2 \hat{s}} \;,
\end{eqnarray}
where $g^{}_{\rm V} = 1/2 - 4 \sin^2{\theta^{}_{\rm w}}/3$ and $g^{}_{\rm A} = 1/2$ for ${q} = {U}$, and $g^{}_{\rm V} = -1/2 + 2\sin^2{\theta^{}_{\rm w}}/3$ and $g^{}_{\rm A} = -1/2$ for ${q} = { D}$, with $\sin^2{\theta^{}_{\rm w}} \approx 0.231$ being the weak mixing angle. The rest of the derivation is similar to the charged-current case.

\section{Leptoquark Models}
The relevant Yukawa interactions of $S_1$ LQ with matter is as follows:
\begin{equation}
\begin{aligned}
\mathcal{L}_{S_1} &\supset  y_{1 i j}^{\rm L L} \overline{Q_{\rm L}^{{\rm c} i, a}} S_{1} \epsilon^{a b} L_{\rm L}^{j, b}+y_{1 i j}^{\rm R R} \overline{U_{\rm R}^{{\rm c} i}} S_{1} E_{\rm R}^{j} +\mathrm{h.c.} \\&  \equiv -y_{1i j}^{\rm L L} \overline{D_{\rm L}^{\rm c i}} S_{1} \nu_{\rm L}^{j}+\left(V^{\rm T} y_{1}^{\rm L L}\right)_{i j} \overline{{U}_{\rm L}^{{\rm c}i}} S_{1} E_{\rm L}^{j} +y_{1 i j}^{\rm R R} \overline{U_{\rm R}^{{\rm c}i}} S_{1} E_{\rm R}^{j} +\mathrm{h.c.},
\end{aligned}
\end{equation}
where $V$ represents the CKM mixing matrix, $y_{1 i j}$ denote the elements of an arbitrary complex $3 \times 3$ Yukawa matrix, and the flavor and  $SU(2)$ indices are denoted by $i,j=1,2,3$ and  $a,b=1,2$, respectively. Here we set the Yukawa texture such a way that $S_1$ LQ can dominantly decay via the following modes: $ S_1 \to \tau {\rm c}, \nu_\tau {\rm s}$. In this way, the bounds on the LQ mass will become less stringent, i.e., $M^{}_{\rm LQ} \gtrsim 400~{\rm GeV}$, if the branching ratios to $\tau {\rm c}$ and $\nu_\tau {\rm s}$ are set to $80\%$ and $20\%$, respectively.

Similarly, for $S_3$ LQ, the relevant Yukawa Lagrangian can be written as:
\begin{equation}
\begin{aligned}
\mathcal{L}_{S_3} &\supset y_{3 i j}^{\rm L L} \overline{Q_{\rm L}^{{\rm c} i, a}} \epsilon^{a b}\left(\tau^{k} S_{3}^{k}\right)^{b c} L_{\rm L}^{j, c}+\text { h.c. }
\equiv -\left(y_{3}^{\rm L L} \right)_{i j} \overline{D_{\rm L}^{\rm c}} S_{3}^{1 / 3} \nu_{\rm L}^{j}-\sqrt{2} y_{3 i j}^{\rm L L} \overline{D_{\rm L}^{\rm c}} i S_{3}^{4 / 3} E_{\rm L}^{j} \\&
+\sqrt{2}\left(V^{\rm T} y_{3}^{\rm L L}\right)_{i j} \overline{U_{\rm L}^{{\rm c} i}} S_{3}^{-2 / 3} \nu_{\rm L}^{j}-\left(V^{\rm T} y_{3}^{\rm L L}\right)_{i j} \overline{U_{\rm L}^{{\rm c} i}} S_{3}^{1 / 3} E_{\rm L}^{j}+ \mathrm{h.c.} ,
\end{aligned}
\end{equation}
where $\tau^k, k=1,2,3$ denotes Pauli matrices. The branching ratios of $S^{}_{3}$ LQ are the same for neutrino and charged lepton final states, which implies a lower limit $M^{}_{\rm LQ} \gtrsim 650~{\rm GeV}$ from the searches of LQ pair production.
For $R_2$ LQ, the relevant part of the Yukawa Lagrangian can be expressed as:
\begin{align}
\mathcal{L}_{R_2} &\supset-y_{2 i j}^{\rm R L} \overline{U_{\rm R}^{i}} R_{2}^{a} \epsilon^{a b} L_{\rm L}^{j, b}+y_{2 i j}^{\rm L R} \overline{E_{\rm R}^{i}} R_{2}^{a *} Q_{\rm L}^{j, a}+\text{h.c.}
\\& 
\equiv -y_{2 i j}^{\rm R L} \overline{U_{\rm R}^{i}} E_{\rm L}^{j} R_{2}^{5 / 3}+\left(y_{2}^{\rm R L} U\right)_{i j} \overline{U_{\rm R}^{i}} \nu_{\rm L}^{j} R_{2}^{2 / 3}
+\left(y_{2}^{\rm L R} V^{\dagger}\right)_{i j} \overline{E_{\rm R}^{i}} U_{\rm L}^{j} R_{2}^{5 / 3 *}+y_{2 i j}^{\rm L R} \overline{E_{\rm R}^{i}} D_{\rm L}^{j} R_{2}^{2 / 3 *}+\mathrm{h.c.}  \notag
\end{align}
There are two states of $R_2$ LQ, one with electric charge $Q=2/3~ (R_{2}^{2 / 3})$ and other one with $Q=5/3 ~(R_{2}^{5 / 3})$. Since $R_{2}^{2 / 3}$ state interacts with neutrino and up-type quark as well as charged lepton and down-type quark, it provides a significant imprint at  tau neutrino telescopes. For $R_2$ LQ,  the Yukawa structure is chosen in such a way that   the state $R_{2}^{2 / 3}$ from $R_2$ LQ can dominantly decay to the following modes: $ R_{2}^{2 / 3} \to \tau {\rm s}, \nu_\tau {\rm c}$. 
For $\tilde{R}_2$ LQ, the relevant Lagrangian is given by
\begin{equation}
\mathcal{L}_{\tilde{R}_2} \supset-\tilde{y}_{2 i j}^{\rm R L} \overline{D_{\rm R}^{i}} \tilde{R}_{2}^{a} \epsilon^{a b} L_{\rm L}^{j, b}+\text{ h.c. } \equiv -\tilde{y}_{2 i j}^{\rm R L} \overline{D_{\rm R}^{i}} E_{\rm L}^{j} \tilde{R}_{2}^{2 / 3}+\left(\tilde{y}_{2}^{\rm R L} \right)_{i j} \overline{D_{\rm R}^{i}} \nu_{\rm L}^{j} \tilde{R}_{2}^{-1 / 3}+\text { h.c. }
\end{equation}
$\tilde{R}_2$ LQ also comprises of two states with electric charges $Q=2/3$ and $-1/3$. However, we find that due to the Yukawa structure, $\tilde{R}_{2}^{-1 / 3}$  coupled to neutrinos dominantly decays to  $ \tilde{R}_{2}^{-1 / 3} \to \nu_\tau$d, $\nu_\tau$s, which requires a LQ mass $1 $ TeV at least to satisfy existing collider constraints.

\bibliographystyle{utcaps_mod}

{\footnotesize
\bibliography{reference}}

\end{document}